\documentclass[aps,prl,reprint,groupedaddress]{revtex4-1}

\usepackage{amsmath,amssymb,amsfonts,bbm,graphicx,times}
\usepackage[dvipsnames]{xcolor}
\usepackage{braket}
\usepackage[normalem]{ulem}
\usepackage[colorlinks=true,bookmarks=false,linkcolor=RoyalBlue,urlcolor=RoyalBlue,citecolor=RoyalBlue,breaklinks]{hyperref}
\usepackage[caption=false, labelfont=bf, labelformat=simple]{subfig}
\usepackage[capitalize]{cleveref}
\newcommand{\eref}[1]{(\ref{#1})}

\renewcommand{\phi}{\varphi}
\renewcommand{\epsilon}{\varepsilon}
\renewcommand{\rho}{\varrho}

\renewcommand{\d}{\mathrm{d}}

\usepackage{units}

\begin{document}
\title{Fingerprint and universal Markovian closure of structured bosonic environments}
\author{Alexander N{\"u}{\ss}eler$^1$, Dario Tamascelli$^{1,2}$, Andrea Smirne$^{2,3}$, James Lim$^{1}$, Susana~F. Huelga$^1$, and Martin~B. Plenio$^{1}$}
\affiliation{$^1$ Institut f{\"u}r Theoretische Physik and IQST,
Albert-Einstein-Allee 11, Universit{\"a}t Ulm, 89069 Ulm, Germany}
\affiliation{$^2$ Dipartimento di Fisica ``Aldo Pontremoli'', Universit{\`a} degli Studi di Milano, Via Celoria 16, 20133 Milano-Italy}
\affiliation{$^3$ Istituto Nazionale di Fisica Nucleare, Sezione di Milano, Via Celoria 16, 20133 Milano-Italy}

%
\begin{abstract}
We exploit the properties of chain mapping transformations of bosonic environments to identify a finite collection of  modes able to capture the characteristic features, or fingerprint, of the environment. Moreover we show  that the countable infinity of residual bath modes can be replaced by a universal Markovian closure, namely a small collection of damped modes undergoing a Lindblad-type dynamics whose parametrization is independent of the spectral density under consideration. We show that the Markovian closure provides a quadratic speed-up with respect to standard chain mapping techniques and makes the memory requirement independent of the simulation time, while preserving all the information on the fingerprint modes. We illustrate the application of the Markovian closure to the computation of linear spectra but also to non-linear spectral response, a relevant experimentally accessible many body coherence witness for which efficient numerically exact calculations in realistic environments are currently lacking.  
\end{abstract}
\maketitle
%


Much theoretical research in recent decades has focused on the study of open quantum systems (OQS) interacting with structured non-Markovian environments~\citep{rivas14,breuer16,deVega17,wiseman18}. Analytical results are hard to obtain except for very specific models, and numerical simulations are typically very challenging unless severe approximations are made. Despite these challenges, the interaction of an OQS with the surrounding environment is not only unavoidable in practice, but there are important instances where non-Markovianity may be instrumental for different manifestations of the presence of quantum coherence or quantum correlations to persist for significant time scales \cite{prior10,christensson2012origin,tiwari2013electronic,chin13,tomasi2021environmentally} or even in the thermodynamical limit \cite{huelga2012non,heineken2021quantum,ask2022non}. 

For thermal bosonic environments, numerical methods developed for a general treatment of non-Markovian problems include e.g.\ Hierarchical Equations of Motion (HEOM)~\citep{tanimura89,tanimura90}, Quasi-Adiabatic Path Integrals (QUAPI)~\citep{makri92, makri93}, Nonequilibrium Green's Function (NEGF) techniques~\citep{rammer86, daniel84}, Non-Markovian Quantum State Diffusion (NMQSD) and similar stochastic methods~\citep{diosi97,diosi14,piilo08} 
and the Time-Evolving Density Operator with Orthogonal Polynomials Algorithm (TEDOPA)~\cite{prior10,chin10,tama15,Woods2015,kohn18,tama19,tama20Entro}. {Recently, powerful hybrid methods have been developed 
including the Time-Evolving Matrix Product Operators (TEMPO)~\citep{Strathearn2018}, merging path integral and tensor network methods, the 
Dissipation-Assisted Matrix Product Factorization (DAMPF)~\cite{somoza19,somoza22}, combining tensor networks and local Markovian dissipators and methodologies that utilize tensor networks in combination with quantum state diffusion \cite{flannigan2022many}.} 
In this work we will develop an exact hybrid scheme that combines the intuition behind strategies aimed at redefining the system-environment boundary by means, for instance, of introducing a reaction coordinate \cite{strasberg2016nonequilibrium} or utilizing surrogate oscillator modes \cite{mascherpa20} with chain mapping transformations as those implemented by TEDOPA \cite{chin10}.

TEDOPA is a certifiable and numerically exact method to efficiently treat OQS dynamics. It first maps the continuum of bath modes unitarily onto a one-dimensional chain of harmonic oscillators, and then it exploits Time-Dependent Density Matrix Renormalization Group (tDMRG)~\cite{schollwock11} to efficiently simulate the dynamics of the resulting configuration. Beside providing an optimal discretization of the bath \cite{devega15discr}, TEDOPA treats the OQS and the bath degrees of freedom on the same footing, thus leaving the possibility of inspecting the evolution of both.
{Moreover, unlike some of the methods mentioned in the previous paragraph, TEDOPA is not restricted to thermal or Gaussian initial states of the environment. However, despite its usefulness TEDOPA, as all methods mentioned earlier, remains computationally intensive. In order to address more challenging questions, such as the determination of multi-dimensional spectra~\cite{JonasARPC2003,BrixnerJCP2004}, further efficiency gains are required. }
\begin{figure}
    \centering
    \includegraphics[width = 1.\columnwidth]{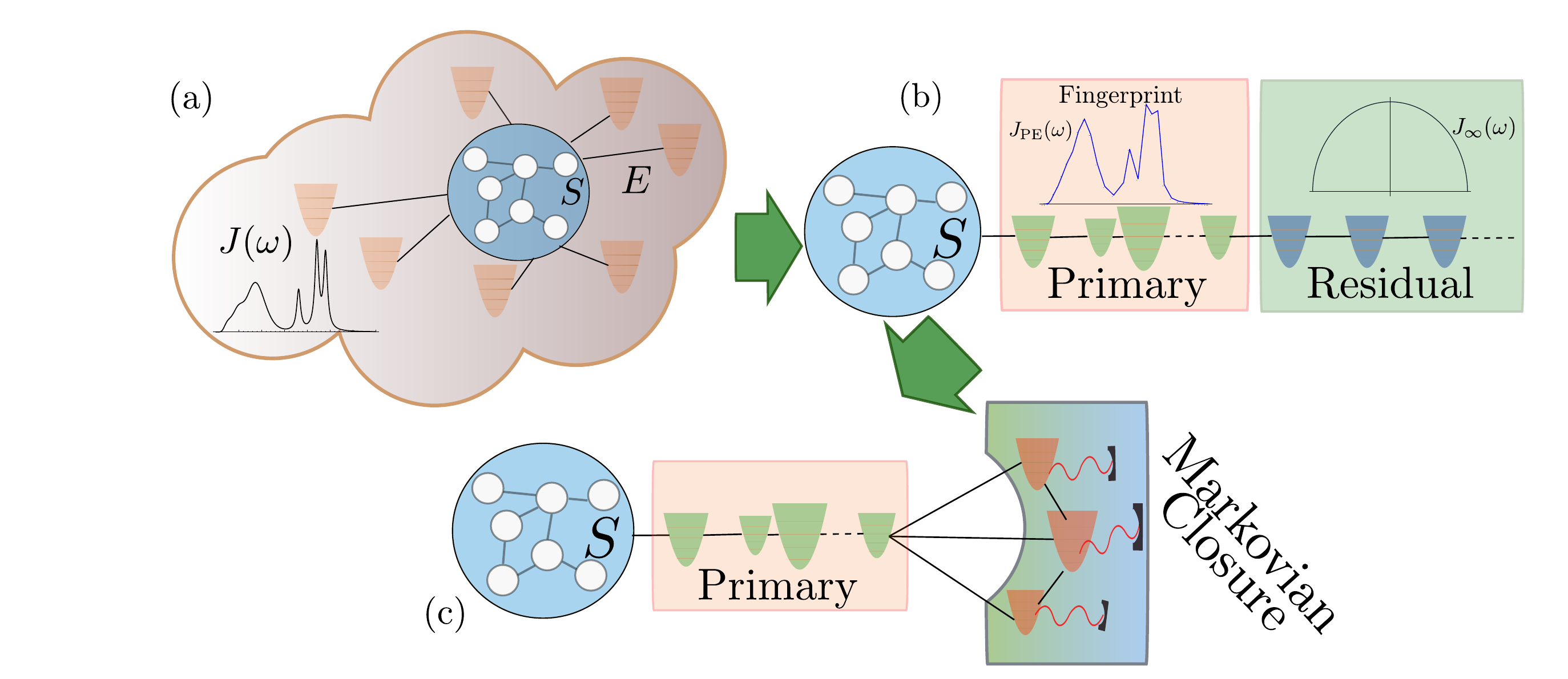}
    \caption{\textbf{Schematics of our procedure:} (a) The  system (S) interacting with a bosonic environment (E), as described by Eqs.1-4. (b) After the chain-mapping, the system interacts with the primary environment, which interacts in turn with the residual environment. (c) The residual environment is replaced by a finite set of interacting damped modes, i.e. the Markovian closure.}
    \label{fig:my_label}
\end{figure}

We will exploit the properties of the TEDOPA mapping to show that the main features, or {fingerprint}, of the environment are captured by a primary bath comprising a finite, and typically small, number of modes directly interacting with the OQS. Moreover, we show that the remaining environmental modes form a universal residual bath and that such a residual bath can be replaced by a finite number of damped harmonic modes undergoing a Lindblad-type dynamics. We provide an explicit construction of this universal Markovian closure and illustrate that the ensuing reduction of the computational resources {renders the calculation of 2D nonlinear electronic spectra~\cite{JonasARPC2003,BrixnerJCP2004} accessible to this method.}



\textit{TEDOPA mapping.} 
We consider a system interacting with a bosonic environment. The complete  Hamiltonian reads
\begin{align}
    H & =\label{eq:totHam} 
H_S + H_E+H_I \\    
H_E &=  \int \d \omega \;   \omega a_\omega^\dagger a_\omega \label{eq:freeHam} \\
    H_I &=  \int \d \omega \;  h(\omega) (A_S^\dagger a_\omega+A_S a_\omega^\dagger),
    \label{eq:initial_interaction_hamiltonian}
\end{align}
($\hbar = 1$) where $H_S$ is the (arbitrary) free system Hamiltonian, $H_E$ describes the free evolution of the bosonic environmental degrees of freedom, and $H_I$  is the bilinear system-environment interaction Hamiltonian \cite{leggett87}. Moreover we assume that $h(\omega)$ has finite support $[\omega_\text{min},\omega_\text{max}]$, $\omega_\text{min} <\omega_\text{max}$, and introduce the spectral density
\begin{equation}
    J(\omega) = \pi h^2(\omega).
    \label{eq:sd}
\end{equation}
As shown in \cite{chin10,prior10,woods14} the Hamiltonian \eref{eq:totHam} can be unitarily mapped to an equivalent one describing
a countably infinite set of modes with operators $b_n^{(\dagger)}$
satisfying the bosonic commutation relations 
$[b_n,b_m^\dagger]=\delta_{nm}$ yielding 
\begin{align} \label{eq:chainHam}
    \begin{split}
        H^C &= H_S + H_I^C+H_E^C, \\
        H_I^C &= \kappa_0 (A_S^\dagger b_1+A_S b_1^\dagger ), \\
        H_E^C & = \sum_{n=1}^{\infty} \omega_n b_n^\dagger b_n +  \kappa_n(
        b_{n+1}^\dagger b_n + b_n^\dagger b_{n+1}),
    \end{split}
\end{align}
where the coefficients $\omega_n$ and $\kappa_n$ are determined either analytically \cite{chin10} or numerically 
~\cite{gautschi94,gautschi04}.

For the following, the joint initial state of system and environment is assumed to factorize, i.e. $\rho_{SE}(0) = \rho_S(0) \otimes \rho_{E}(0)$, with $\rho_{E}(0)$ a thermal state. 
Following \cite{tama19}, it is possible to replace $\rho_E(0)$ with the factorized pure vacuum state when performing at the same time a suitable transformation of the spectral density.

\textit{Bath fingerprint and universal closure.}
\begin{figure*}[t]
    \centering
    \includegraphics[width=2. \columnwidth]{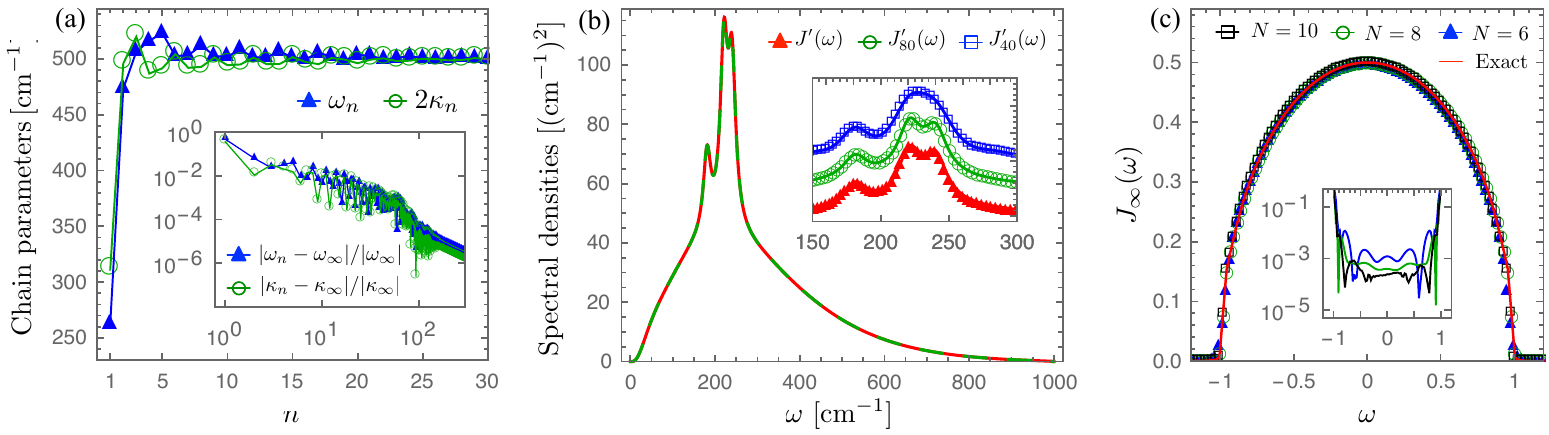}
    \caption{\textbf{Coefficients, Fingerprint and Closure:} 
    (a) The TEDOPA chain oscillator frequencies $\omega_n$ (blue triangles) and couplings $\kappa_n$ (green circles) obtained for the WSCP spectral density $J(\omega)$ described in the main text. 
    In the inset the relative errors between the exact and asymptotic coefficients.
    (b) The {effective spectral density $J'(\omega)$ of the TEDOPA chain truncated after 400 sites (red line)  and the effective spectral density $J_{M}'(\omega)$ with the asymptotic values $\Omega/ K$ replacing the exact coefficients $\omega_n/\kappa_n$ for $n>M=80$ (dashed green line). In the inset, the effective spectral densities $J'(\omega)$, $J_{80}'(\omega)$, $J_{40}'(\omega)$ (vertically shifted for improved visibility). 
    } (c) The residual spectral density \eref{eq:asymptoticResidual} for $[\omega_\text{min},\omega_\text{max}] = [-1,1]$ and the approximation provided by  TSO with $N=6,8,10$ auxiliary modes; inset: the relative error $|J_\infty(\omega) - J_N^\text{TSO}(\omega)|/J_\infty(\omega)$ for $\omega \in [-1,1]$. 
    }
    \label{fig:dimer_spectral_density}
\end{figure*}
As shown in \cite{chin10,woods14},
the asymptotic values of the chain mode frequencies $\omega_n$ and coupling constants $\kappa_n$ are
\begin{align} 
    \omega_n &\xrightarrow[n \to \infty] { } \frac{\omega_\text{min}+\omega_\text{max}}{2} \stackrel{\text{def}}{=} \Omega,  \nonumber \\
    \kappa_n &\xrightarrow[n \to \infty] { } \frac{\omega_\text{max}-\omega_\text{min}}{4} \stackrel{\text{def}}{=} K. \label{eq:limCoeff}
\end{align}
%
%
Given the spectral density $J(\omega)$, for any  $M>0$ we define Hamiltonian of the residual bath as
\begin{align}
    H_R &= \Omega \hspace{-7pt}\sum_{m=M+1}^{\infty} \hspace{-3pt} b_m^\dagger b_m   
+ K \hspace{-7pt}\sum_{m=M+1}^{\infty}
\hspace{-3pt}(b_{m+1}^\dagger b_m + b_m^\dagger b_{m+1}), \label{eq:truncApproxHam}
\end{align}
where $H_R$ is obtained from $H_E^C$ by disregarding the first $M$ modes and replacing the chain coefficients $\omega_m/\kappa_m$ with their asymptotic values.
%
For a given value of $M$ the original system-environment Hamiltonian \eref{eq:chainHam} is approximated by 
\begin{align} \label{eq:appChain}
    &\widetilde{H} = H_S+H_I^C+H_\text{PE}+\kappa_M (b_M^\dagger b_{M+1} + \text{h.c.})+H_R, \\
    &H_\text{PE} = \sum_{n=1}^{M} \omega_n b_{n}^\dagger b_{n} +\sum_{n=1}^{M-1} \kappa_n (b_n^\dagger b_{n+1} + b_n b_{n+1}^\dagger).
\end{align}
Of course, equality $\widetilde{H} = H^C$ holds only in the  $M \to \infty$ limit: for finite values of $M$ only the chain coefficients associated to the first $M$ chain sites are exact, while the remaining ones are only approximated. Such an approximation is, however, under full control: the exact coefficients are known and the effects of small variations of the spectral density on the system dynamics can be bounded analytically~\cite{mascherpa17}. In what follows we denote with $M(\epsilon)$ the smallest $M$ such that  $|(\omega_m-\Omega)/\Omega|,| (\kappa_m-K)/K|< \epsilon,\ \forall m \geq M$; we moreover chose $\epsilon = 10^{-3}$ and denote this case as $\epsilon$-converged, but other choices can be used.
%

 The exact part of the chain, comprising the first $M$ modes, plays the role of \emph{primary environment} (PE), capturing the specific features, or \emph{fingerprint}, of the spectral density.
 %
This is exemplified in Fig.~\ref{fig:dimer_spectral_density} where we consider {an environmental spectral density $J(\omega) = J_\text{AR}(\omega) + \sum_{k=1}^3 J_{L,k}(\omega)$ of a photosynthetic pigment-protein complex, Water-Soluble Chlorophyll-binding Protein (WSCP) from cauliflower \cite{renger15}. 
Here $J_\text{AR}(\omega)$ is a broad phonon spectrum originating from protein motions~\cite{renger15} while three narrow Lorentzian peaks $J_{L,k}(\omega)$ describe intrapigment vibrational modes~\cite{PieperJPCB2011} with vibrational frequencies $(\Omega_1,\Omega_2,\Omega_3) = \unit[(181, 221, 240)]{cm^{-1}}$ and an energy damping rate on the picosecond time scale (see the SI \cite{suppl} for more details).} The domain $[\omega_\text{min}, \omega_{\max}]=[0,1000]\,\text{cm}^{-1}$ of {$J(\omega)$} is chosen such that the discarded weight $W_{d}(\omega_\text{max}) := \int_{\omega_\text{max}}^{\infty} \d \omega^\prime J(\omega^\prime)/\int_0^{\infty} \d \omega^\prime J(\omega^\prime)< 10^{-3}$. As Fig.~\ref{fig:dimer_spectral_density}(a) shows, the chain coefficients $\omega_n, \kappa_n$ are $\epsilon=10^{-3}$-converged for $M\approx 10^2$. 
{In order to better illustrate the meaning of the fingerprinting modes, it is appropriate to truncate the TEDOPA chain after a finite number of sites, thus comparing discrete spectral densities to discrete spectral densities. In Fig.~\ref{fig:dimer_spectral_density}(b), the  effective spectral density $J'(\omega)$ corresponding to a chain truncated after 400 sites is shown in red.
Fig.~\ref{fig:dimer_spectral_density}(b) also shows the effective spectral density $J_{M}'(\omega)$ obtained by replacing the exact coefficients $\omega_n/\kappa_n$ of the truncated TEDOPA chain with their asymptotic values $\Omega/K$ for $n>M$. It is clear that $J_{M}'(\omega)$ is in excellent agreement with the exact $J'(\omega)$ for $M\gtrsim 80$: the three narrow Lorentzian peaks of the WSCP spectral density in the region $\omega \in [150,300]\,\unit[]{cm^{-1}}$ are well resolved for $M=80$, while for $M=40$ the fine structure is lost, as shown in the inset of Fig.~\ref{fig:dimer_spectral_density}(b)} (see also the SI~\cite{suppl}).
The fingerprint of the WSCP spectral density is therefore provided by a PE {consisting of a finite number $M$ of sites}; a suitable value of $M$ can be determined by means of the $\epsilon$-convergence criterion, which in turn determines the number of sites after which the exact semi-infinite chain, comprising the modes with $n>M$, is practically indistinguishable from an approximating \emph{residual environment} governed by $H_R$ (see Eq.~\eref{eq:truncApproxHam}). 

 \textit{Markovian closure.} 
The spectral density of the residual environment reads~\cite{chin10,woods14}
\begin{equation}
    J_\infty(\omega) = K^2\frac{\sqrt{(\omega-\omega_\text{min})(\omega_\text{max}-\omega)}}{2},\label{eq:asymptoticResidual}
\end{equation}
%
%
{also known as Winger semicircle \cite{arnold71,kohn21}}. Since $J_\infty(\omega)$ does not depend on the specific spectral density that we started with, but only on $\omega_\text{min/max}$, the residual environment is universal. Due to the translational invariance of the residual environment, an excitation entering the semi-infinite homogeneous part of the chain propagates ballistically away from the PE \cite{tama20Entro} at a speed proportional to $ K$. {On the one hand, this observation makes it clear that 
to avoid finite-size effects on the dynamics of the system, or of the PE, the truncation of the (semi-infinite) chain, which is required to enable simulations, must be suitably chosen: for a given simulation time $T$ the length of truncated chain must be proportional to $K T$. Long-time simulations can therefore become computationally highly demanding, since a very large number of chain oscillators must be considered. On the other hand, the same ballistic propagation} provides a suggestive picture of the irreversibility of the interaction of the extended system with the residual environment, with the latter ``absorbing '' all the excitations coming from the primary environment. Our findings show that such an absorption mechanism can be realized by means of a finite environment made up of a  small collection of harmonic oscillators governed by the nearest-neighbor coupling Hamiltonian
\begin{align}
    H_{\text{aux}} &:= \sum_{n = 1}^N \Omega_n d_n^\dagger d_n + \sum_{n = 1}^{N-1} g_n (d_{n+1}^\dagger d_{n} + d_{n}^\dagger d_{n+1})
    \label{eq:tso_hamiltonian}
\end{align}
and subject to the local Lindblad dissipator
\begin{align}
    \mathcal{D}(\rho) := \sum_{n = 1}^{N} \Gamma_n \left(d_n \rho d_n^\dagger - \frac{1}{2} \{d_n^\dagger d_n, \rho\} \right),
    \label{eq:tso_dissipator}
\end{align}
where $\rho$ is a statistical operator living on the Hilbert space of the surrogate oscillators providing a \emph{Markovian Closure} (MC) of the chain. For given $N$, the coefficients $\Gamma_n, \Omega_n, g_n$  are determined by means of the  Transformation to Surrogate Oscillators (TSO) procedure \cite{tamascelli18,mascherpa20}.
It is important to stress here that the universality of the residual spectral density $J_\infty$ implies that the derivation of the TSO coefficients must be done only once. Figure \ref{fig:dimer_spectral_density}(c) shows the spectral density of the auxiliary system provided by TSO for the choices $N=6,8,10$, and allows one to appreciate the accuracy of the approximation of $J_\infty$. Full detail on the TSO derivation and the values of the TSO parameters for different closure sizes are provided in the SI \cite{suppl}. 
The use of MC is most effective when long-time dynamics is considered. In this case, in fact, standard TEDOPA would require the use of very long chains as not to produce finite-size effects affecting the system and the extended system, comprising the system and the PE, dynamics.  In what follows we describe a relevant application of the MC. 


%

\begin{figure}
    \centering
    \includegraphics[width=1\linewidth]{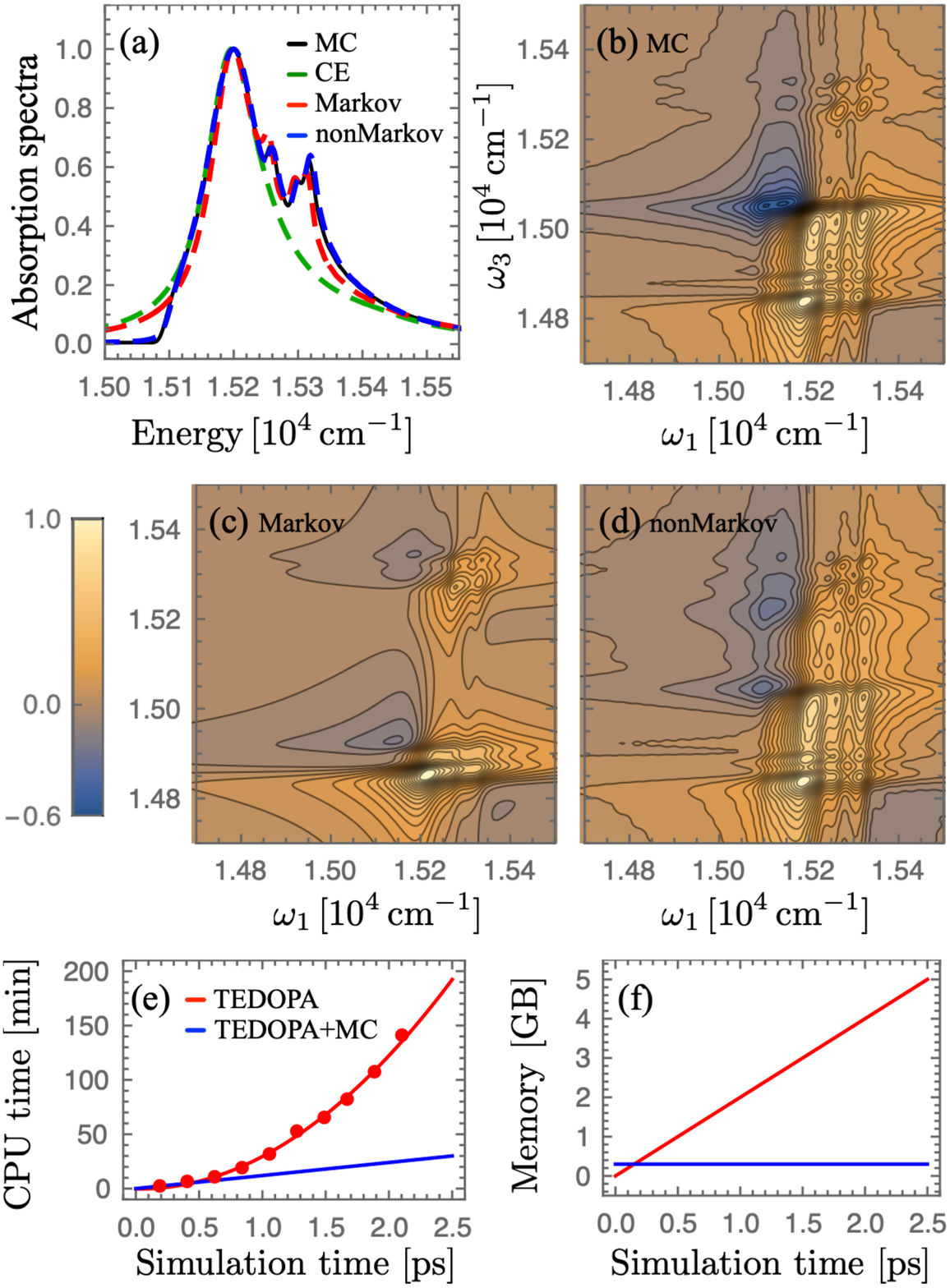}
    \caption{\textbf{Linear and 2DES optical spectra of WSCP, computational cost:} (a) Absorption spectra computed by TEDOPA with Markovian closure (MC), cumulant expansion (CE), reduced vibronic models with Markovian and non-Markovian effects of a broad environmental spectrum $J_{\rm AR}(\omega)$. 2D electronic spectra at $t_2=500\,{\rm fs}$ computed by (b) MC, and reduced vibronic models with (c) Markovian and (d) non-Markovian effects of the broad noise spectrum. The time (e) and memory scaling (f) as functions of the simulation time $t_\text{max}$.}
    \label{fig:2DES}
\end{figure}

\textit{Application: long-time system dynamics.}
To demonstrate that our construction allows for numerically exact long-time simulations, we investigate optical responses of a model molecular complex, WSCP, an aggregate for which a comprehensive spectral charaterization is available ~\cite{renger15}. The dimeric system consists of two interacting pigments, with each pigment coupled to a local phonon environment characterized by $J(\omega)$. Motivated by the actual parameters of WSCP, we model each pigment as a two-level system with identical energy-gap between electronic ground and excited states ($\sim 15198\,{\rm cm}^{-1}$) and consider an electronic coupling $69\,{\rm cm}^{-1}$ between pigments (see the SI \cite{suppl}). As to make most of the details of the spectra clearly visible, we consider a phonon bath at zero temperature and parallel transition dipoles of pigments. In this setting the relevant state space is spanned by a low-energy optically dark electronic eigenstate (or exciton) $\ket{E_d}$, and a bright high-energy exciton state $\ket{E_b}$, which dominates the optical responses. As shown in Fig.\ref{fig:2DES}(a), the (numerically exact) absorption spectrum computed by the MC cannot be reproduced by a line shape theory based on second order cumulant expansion~\cite{NovoderezhkinJPCB2004,renger2006,AbramaviciusJCP2010}, which has been widely used to simulate optical responses of photosynthetic systems. The absorption spectrum can be better reproduced by a reduced vibronic model where the three narrow peaks $J_{L,k}(\omega)$ in the spectral density are included in the system Hamiltonian in addition to electronic states, and the remaining broad environmental spectrum $J_{\rm AR}(\omega)$ is considered a source of Markovian noise~\cite{LimNC2015,Felipe2021}. However, the low-energy part of the absorption spectrum cannot be reproduced. The numerically exact absorption spectrum can be well reproduced only when multiple narrow Lorentzian functions are fitted to the spectral density $J_{\rm AR}(\omega)$ and each Lorentzian is considered a damped harmonic oscillator coupled to electronic states~\cite{tamascelli18}. Such a non-Markovian treatment of the broad environmental spectrum $J_{\rm AR}(\omega)$ enables one to successfully reproduce numerically exact absorption line shape, as shown in Fig.\ref{fig:2DES}(a). For more details of the reduced models and the origin of absorption peaks, see the SI~\cite{suppl}.

The non-Markovian treatment of the broad environmental spectrum becomes more important when long-time system dynamics is considered. As an example, we consider two-dimensional electronic spectroscopy (2DES)~\cite{JonasARPC2003,BrixnerJCP2004}, which is a time-resolved optical technique for measuring electronic and vibrational dynamics~\cite{LimNC2015,PlenioJCP20132DES,RomeroNP2014,FullerNC2014,LimPRL2019}. In 2DES, a sample is perturbed by three laser pulses with controlled time delays, enabling one to investigate molecular dynamics as a function of excitation $\omega_1$ and detection energies $\omega_3$ for each time delay $t_2$ between the second and the third pulse. Fig.\ref{fig:2DES}(b) shows numerically exact stimulated emission component of rephasing 2D spectra
at a waiting time $t_2=500\,{\rm fs}$ (see the SI~\cite{suppl}), where multiple 2D peaks are present for $\omega_3<\omega_1$. These results cannot be reproduced by a reduced vibronic model when the broad phonon spectrum $J_{\rm AR}(\omega)$ is considered a source of the Markovian noise, as shown in Fig.\ref{fig:2DES}(c). In order to better reproduce the numerically exact nonlinear optical responses, it is necessary to include the non-Markovian effects of the broad noise spectrum in the reduced model (see Fig.\ref{fig:2DES}(d)). These non-Markovian features in 2DES originate from multi-phonon transitions where initial bright exciton state $\ket{E_b,0}$ is vibronically mixed with lower-energy dark exciton states $\ket{E_d,1_j}$ while creating a single vibrational excitation in the $j$-th mode, and then goes back to the bright exciton states $\ket{E_b,1_j,1_k}$ while creating an additional vibrational excitation in the $k$-th mode.
As the single-phonon transitions result in optically dark states $\ket{E_d,1_k}$, the multi-phonon transitions can dominate 2D line shape when the system has enough time to evolve from the initial vibrational ground state $\ket{E_b,0}$ to the final doubly-excited vibrational states $\ket{E_b,1_j,1_k}$, as is the case of a long waiting time $t_2=500\,{\rm fs}$ considered here. These results could be most relevant in the interpretation of optical responses of charge-separating systems, such as reaction centers in photosynthetic systems~\cite{RomeroNP2014,FullerNC2014} and organic solar cells~\cite{DeSioNC2016}, where light-absorbing bright excitons relax to optically dark charge-transfer states where electrons and holes are spatially separated while creating vibrational excitations.

{{\em Performance --}} In order to allow for a comparison with the results obtained by a standard TEDOPA implementation, we limited our investigation to waiting time $t_2 = \unit[500]{fs}$, resulting in an overall simulation time of up to \unit[2.5]{ps}. For such evolution times, converged TEDOPA simulations required about $N=240$ chain sites to avoid any finite size effects, whereas with the MC only $M+6=86$ sites were needed for $\epsilon = 10^{-3}$ convergence of the chain coefficients. This translates into a significant reduction of the computational cost: the computation time is reduced from $\approx\unit[190]{min}$ to $\approx\unit[30]{min}$ (using 12 Intel Xeon Cascade Lake cores) while the memory consumption is reduced from $\approx \unit[5]{GB}$ to $\approx \unit[300]{MB}$. Most importantly, the MC allows for a quadratic speed-up of TEDOPA simulations {for given simulated physical time}. As a matter of fact, for standard TEDOPA the length of the chain increases linearly with the simulation time $t_\text{max}$, resulting in a CPU time-cost $O(t_\text{max}^2)$, whereas for the MC this cost is reduced to $O(t_\text{max})$. For the same reason the memory required by TEDOPA simulations scales linearly with $t_\text{max}$ while the MC the memory requirement is constant, as shown in Fig.~\ref{fig:2DES}(e)-(f).

\textit{Conclusion and outlook.} Besides providing a clear computational advantage when long-time evolution is considered, the MC preserves the possibility offered by TEDOPA to treat the system and the (primary) environment on the same footing. The information on the relevant environmental degrees of freedom, the fingerprint, is therefore fully available for inspection. In this sense, the MC is complementary to standard TEDOPA and provides us with a most powerful tool for the study of unitary equilibration processes in fundamental systems such as the single-impurity Anderson model \cite{anderson61,schwarz18,kohn21,nuessler20} and in the context of quantum thermodynamics~\cite{lewenstein19,tanimura2020}. Moreover, we expect the MC to be able to significantly reduce the computational overhead in situations characterized by long-range correlations between different environments, as in 1D transport models~\cite{kohn18,deVega15}, {and electron scattering problems where the populations of the electron wave packets propagating through semi-infinite electrodes after scattering events are evaluated~\cite{vittmann22}.} The definition of the environmental fingerprint can also be exploited in other simulation techniques, such as HEOM, to simplify the fitting procedure~\cite{tanimura2020,Felipe2021}. In future work we will exploit the MC for the determination {of} higher order multi-phonon transitions and the computation of the non-linear spectral response of aggregates of biological relevance as well as for the study of energy and charge transfer in general many body systems and specifically light harvesting aggregates of both natural and artificial nature.

{\em Acknowledgements --} This work was supported by the DFG via QuantERA project ExtraQt and the ERC Synergy grant HyperQ, and support by the state of Baden-Württemberg through bwHPC and the German Research Foundation (DFG) through Grant No INST 40/575-1 FUGG (JUSTUS 2 cluster).

\nocite{satoh01,
renger15,
PieperJPCB2011,
Lim_NJP2014,
Felipe2021,
Mukamel1995,
NovoderezhkinJPCB2004,
renger2006,
AbramaviciusJCP2010,
tamascelli18,
May2000,
breuer02,
RengerJPP2011,
MuhBA2012,
Prep,
LimNC2015,
JonasARPC2003,
BrixnerJCP2004,
LimPRL2019,
mascherpa20,
beylkin2005,
garraway97,
May2011,
Hornberger2009,
lubich15,
haege16}

\bibliography{MarkovianClosure}

\begin{thebibliography}{27}
\bibitem [{con()}]{conv}%
  \BibitemOpen
  \href@noop {} {}\bibinfo {note} {The local dimension is determined by means
  of numerical convergence: it is increased until the difference of the average
  occupation number of the chain oscillators with $D=d$ or $D=d+1$ levels falls
  below some threshold (in our case $10^{-12}$). The same procedure is applied
  for the determination of the bond dimension $\chi_{ \left | {\psi} \right
  \rangle}$.}\BibitemShut {Stop}%
\end{thebibliography}

\end{document}


\title{Supplemental Material: Fingerprint and Universal Markovian closure of structured bosonic environments}
\author{Alexander N{\"u}{\ss}eler$^1$, Dario Tamascelli$^{1,2}$, Andrea Smirne$^{2,3}$, James Lim$^{1}$, Susana~F. Huelga$^1$, and Martin~B. Plenio$^{1}$}
\affiliation{$^1$ Institut f{\"u}r Theoretische Physik and Center for Integrated Quantum Science and Technology (IQST),\\
Albert-Einstein-Allee 11, Universit{\"a}t Ulm, 89069 Ulm, Germany}
\affiliation{$^2$ Dipartimento di Fisica ``Aldo Pontremoli'', Universit{\`a} degli Studi di Milano, Via Celoria 16, 20133 Milano-Italy}
\affiliation{$^3$ Istituto Nazionale di Fisica Nucleare, Sezione di Milano, Via Celoria 16, 20133 Milano-Italy}

%
\maketitle
%


The structure of this Supplemental Information is as follows: in Section \ref{sec:WSCPmodel}  we provide more detail on the WSCP dimer. Sections \ref{sec:reduced} and \ref{sec:2d} are devoted to the computation of the corresponding absorption and two-dimensional electronic  spectra (2DES). 
In Section~\ref{sec:resenv} we quantify the error introduced by the residual environment compared to the exact solution while, in Section \ref{sec:TSO}, we describe in detail the procedure for the derivation of the Markovian Closure parameters. Section \ref{sec:twm} presents a comparison between standard TEDOPA and TEDOPA with Markovian Closure, assessing the quality of the results and the computational advantage provided by the Markovian Closure technique.

\section{WSCP} \label{sec:WSCPmodel}

Water-Soluble Chlorophyll Protein (WSCP) of Cauliflower ({\it Brassica oleracea L.})~\cite{satoh01} is a dimeric pigment-protein photosyntetic complex.  The dimer consists of two interacting pigments, with each pigment coupled to a local phonon environment, originating from  the motion of the proteins scaffolding the pigments~\cite{renger15}, and from intrapigment vibrations~\cite{PieperJPCB2011}.

We thus model WSCP as an excitonically coupled dimer where the electronic excitation of each site is
coupled to a local vibrational environment characterized by a phonon spectral density $J(\omega)$. The total Hamiltonian consists of three parts, $H=H_{e}+H_{v}+H_{e-v}$.
The electronic Hamiltonian $H_{e}$ is characterized by on-site energies $E_i$ and inter-site electronic
coupling $V$
\begin{equation}
	H_{e}=\sum_{j=1}^{2}E_j \ketbra{e_j}{e_j} +
    V(\ketbra{e_1}{e_2}+\ketbra{e_2}{e_1}),
\end{equation}
where $\ket{e_j}$ represents an electronic state where site $j$ is excited and the other site is in its ground state. Such a single electronic excitation is transferred between two sites mediated by the electronic coupling $V$. In this work, we consider identical on-site energies, $E\equiv E_1=E_2\approx 15198\,{\rm cm}^{-1}$, corresponding to 658\,nm, for which the electronic eigenstates (excitons) are fully delocalized over two sites, $\ket{E_b}=\frac{1}{\sqrt{2}}(\ket{e_1}+\ket{e_2})$ and $\ket{E_d}=\frac{1}{\sqrt{2}}(-\ket{e_1}+\ket{e_2})$. We consider the electronic coupling $V=69\,{\rm cm}^{-1}$ of WSCP homodimers~\cite{renger15}. The vibrational environments are modelled by quantum harmonic oscillators
\begin{equation}
	H_{v}=\sum_{j=1}^{2}\sum_k\omega_k b_{j,k}^{\dagger} b_{j,k},
\end{equation}
where $b_{j,k}$ describes a vibrational mode with frequency $\omega_k$, which is locally coupled to site $j$.
The coupling of the electronic degrees of freedom to their respective vibrational environment is described
by
\begin{equation}
	H_{e-v}=\sum_{j=1}^{2}\ket{e_j}\bra{e_j}\sum_{k}\omega_{k}\sqrt{s_{k}}(b_{j,k}+b_{j,k}^{\dagger}),
\end{equation}
where the vibronic coupling strength is quantified by a Huang-Rhys factor $s_{k}$. In this representation, the spectral density is defined by $J(\omega)=\sum_{k}\omega_k^2 s_k \delta(\omega-\omega_k)$ with $\delta(x)$ denoting the Dirac delta function.

To identify the influence of vibrational modes on electronic dynamics and associated optical responses, we define the center-of-mass $B_{k}=\frac{1}{\sqrt{2}}(b_{1,k}+b_{2,k})$ and relative motion modes $b_{k}=\frac{1}{\sqrt{2}}(b_{1,k}-b_{2,k})$ of the local vibrational modes $b_{1,k}$ and $b_{2,k}$ with
identical frequency $\omega_k$~\cite{Lim_NJP2014,Felipe2021}. $B_k$ and $b_k$ satisfy the bosonic commutation relations and the total Hamiltonian can
be expressed as $H=H_{e}+H_{c}+H_{r}$, where
\begin{align}
	H_{c}&=\sum_{k}\omega_{k}B_{k}^{\dagger}B_{k}+(\ket{e_1}\bra{e_1}+\ket{e_2}\bra{e_2})\sum_{k}\omega_{k}\sqrt{\frac{s_{k}}{2}}(B_{k}+B_{k}^{\dagger}),\label{eq:lim_H_c}\\
	H_{r}&=\sum_{k}\omega_{k}b_{k}^{\dagger}b_{k}+(\ket{e_1}\bra{e_1}-\ket{e_2}\bra{e_2})\sum_{k}\omega_{k}\sqrt{\frac{s_{k}}{2}}(b_{k}+b_{k}^{\dagger}).
\end{align}
Note that the center-of-mass modes $B_k$ are coupled to both electronic excited states $\ket{e_1}$ and
$\ket{e_2}$ with the same coupling strength and phase. This implies that the coupling to the
center-of-mass motion, described by $H_c$, does not affect electronic dynamics within the single excitation
manifold. Instead, it induces dephasing of optical coherences between electronic ground and excited states,
which broadens the line shapes of linear and nonlinear optical spectra. The contribution of the center-of-mass modes to absorption and two-dimensional electronic spectra can be taken into account analytically~\cite{Felipe2021}, as discussed below in detail.

The vibronic coupling to the relative motion of vibrational modes, described by $H_r$, affects electronic dynamics
in the single excitation subspace. Here we describe $H_r$ in terms of exciton states $\ket{E_b}$ and $\ket{E_d}$ that
diagonalize the electronic Hamiltonian $H_{e}=\sum_{j=b,d}E_j\ket{E_j}\bra{E_j}$ with exciton energies $E_b=E+V$ and $E_d=E-V$. In the exciton basis, $H_r$ is expressed as
\begin{align}
	H_{r}=\sum_{k}\omega_{k}b_{k}^{\dagger}b_{k}-(\ket{E_b}\bra{E_d}+\ket{E_d}\bra{E_b})\sum_{k}\omega_{k}\sqrt{\frac{s_k}{2}}(b_{k}^{\dagger}+b_{k}),
\end{align}
where off-diagonal vibronic (electronic-vibrational) couplings $\ket{E_b}\bra{E_d}(b_{k}^{\dagger}+b_{k})$ induce the transitions between exciton states mediated by the creation and annihilation of vibrational excitations, leading to a vibronic mixing~\cite{LimNC2015,Felipe2021}. These off-diagonal vibronic couplings are perturbatively treated in the second order cumulant expansion and the reduced vibronic model where a broad phonon spectrum $J_{\rm AR}(\omega)$ is approximately described by a Lindblad equation (Markovian noise). The off-diagonal couplings can be treated in a non-perturbative way by using numerically exact methods, such as TEDOPA with Markovian closure considered in this work.

Following \cite{renger15,PieperJPCB2011}, we model the environment as a continuum of bosonic modes and define the spectral density of the WSCP dimer as the sum of two contributions,
\begin{align} \label{eq:contWSCP}
    J(\omega) := J_{\text{AR}}(\omega) + J_{\text{AL}}(\omega).
\end{align}
The first contribution is given by the Adolphs-Renger spectral density,
\begin{align} \label{eq:arsd}
    J_{\text{AR}}(\omega) &:= \sum_{k = 1,2} \frac{\pi c_k}{9!\;2} \, \frac{\omega^5}{\omega_{c,k}^4} \, \e^{-\sqrt{\omega/\omega_{c,k}}},
\end{align}
with weights and cut-off frequencies set to $(c_1,c_2)= (35.45,22.15)$ and $(\omega_{c,1},\omega_{c,2}) = (0.557,1.936) \text{cm}^{-1}$, respectively.
The second contribution comprises three sharp Lorentzian peaks,
\begin{align}
    J_{\text{AL}} (\omega)&:= \sum_{k = 1}^{3} \frac{8S_k \, \Gamma_k \Omega_k (4\Omega_k^2 + \Gamma_k^2) \; \omega}{(4(\omega - \Omega_k)^2 + \Gamma_k^2)(4(\omega + \Omega_k)^2 + \Gamma_k^2)},
\end{align}
 centered at $(\Omega_1,\Omega_2,\Omega_3) = \unit[(181, 221, 240)]{cm^{-1}}$ with Huang-Rhys factors $(S_1,S_2,S_3) = (0.0173, 0.0246, 0.0182)$, and  equal width $\Gamma_k = \Gamma = \unit[20]{cm^{-1}}$, corresponding to a lifetime of $\unit[500]{fs}$.

For this choice of parameters the exact reorganization energies evaluate to
\begin{align}
    \begin{split}
        \lambda_{\text{AR}} &= \lim_{\omega_{\max} \rightarrow \infty} \lambda_{\text{AR}}^{\omega_{\max}} = \int_{0}^{\infty}  \frac{J_{\text{AR}}(\omega)}{\pi\omega} \; \d\omega \approx \unit[62.63]{cm^{-1}} \\
        \lambda_{AL} &= \lim_{\omega_{\max} \rightarrow \infty} \lambda_{\text{AL}}^{\omega_{\max}} = \int_{0}^{\infty}  \frac{J_{\text{AL}}(\omega)}{\pi\omega} \; \d\omega \approx \unit[12.94]{cm^{-1}}.
    \end{split}
    \label{eq:reorganization_energy_wscp}
\end{align}
Note that $\omega_{\max}$ is ultimately defined through \cref{eq:reorganization_energy_wscp} by requesting that the relative error between the exact reorganization energy $\lambda = \lambda_{\text{AR}} + \lambda_{\text{AL}}$ and the approximated reorganization energy $\lambda^{\omega_{\max}} = \lambda^{\omega_{\max}}_{\text{AR}} + \lambda^{\omega_{\max}}_{\text{AL}}$ is below a certain threshold, i.e.,
\begin{align}
    \frac{\lambda - \lambda^{\omega_{\max}}}{\lambda} < \epsilon.
\end{align}
In the example discussed in the main text, we set $\epsilon = 10^{-3}$ which leads to a finite support $[0,\omega_\text{max}]$ of the spectral density with $\omega_{\max} = \unit[1000]{cm^{-1}}$.

We finally observe that, when WSCP spectral density is defined as in Eq.(\ref{eq:contWSCP}), the interaction term $H_{e-v}$ can be rewritten as
\begin{align}
    H_{\text{I}} &:= \sum_{k=1}^{2} \int_{0}^{\omega_{\max}} \d \omega \sqrt{\frac{J(\omega)}{\pi}} \ketbra{e_k} {e_k} (a_{\omega,k} + a_{\omega,k}^\dagger),
    \label{eq:interaction_hamiltonian_wscp}
\end{align}
while the free Hamiltonian of the environment reads 
\begin{align}
    H_{\text{E}} &:= \sum_{k=1}^{2}\int_{0}^{\omega_{\max}} \d\omega \; \omega \; a_{\omega, k}^\dagger a_{\omega, k}.
    \label{eq:vibrational_hamiltonian_wscp}
\end{align}
The renaming $H_e = H_\text{S}$ completes the identification of the total Hamiltonian $H = H_e+H_v+H_{e-v}$ with the Hamiltonian (1) of the main text.


\section{Reduced vibronic models} \label{sec:reduced}

Here we summarize approximate methods for computing linear and nonlinear optical spectra considered in the main text, including second order cumulant expansion~\cite{Mukamel1995,NovoderezhkinJPCB2004,renger2006,AbramaviciusJCP2010}, reduced vibronic models where a broad phonon spectrum $J_{\rm AR}(\omega)$ (see \eref{eq:arsd}) is approximately described by a Lindblad equation~\cite{LimNC2015,Felipe2021} (Markovian noise) or by a set of quantum harmonic oscillators under Lindblad damping~\cite{tamascelli18} (non-Markovian noise).

\subsection{Absorption}

In absorption simulations, we consider an initial state in the form $\ket{g}\bra{g}\otimes \rho_v(T)$ where $\ket{g}$ is a global electronic ground state where both sites are in their electronic ground states, and $\rho_v(T)$ is a thermal state of the vibrational modes at temperature $T$. In this work, we consider zero temperature where each mode is in its vibrational ground state at the initial time; the initial state of the environment is therefore the factorized vacuum state that we indicate by $\rho_v(0) = \ketbra{0}{0}$.

When light-matter interaction is sufficiently weak and can be well described in perturbation theory, the absorption line shape $A(\omega)$ is determined by the Fourier transform of a dipole-dipole correlation function~\cite{Mukamel1995,May2000}
\begin{equation}
	A(\omega)\propto {\rm Re}\left[\omega\int_{0}^{\infty}dt e^{i\omega t}{\rm Tr}[\mu^-\mathcal{U}(t)[\mu^+\ket{g}\bra{g}\otimes \rho_v(T)]]]\right],
	\label{eq:absorption_line_shape}
\end{equation}
where $\mu^+=\sum_{i=1}^{2}(\hat{\epsilon}\cdot\vec{d}_i)\ket{e_i}\bra{g}$ and $\mu^-=\sum_{i=1}^{2}(\hat{\epsilon}\cdot\vec{d}_i)\ket{g}\bra{e_i}$ describe optical transitions between electronic ground and excited states, with $\vec{d}_i$ representing the transition dipole moment of site $i$, and $\hat{\epsilon}$ the polarization of an external electric field inducing absorption process. The time evolution of the electronic-vibrational system perturbed by an optical pulse, described by the application of $\mu^+$, is represented by a unitary operator $\mathcal{U}(t)[A] = e^{-iHt}A e^{iHt}$ with the total Hamiltonian $H$. For simplicity, we consider parallel transition dipole moments, $\vec{d}\equiv \vec{d}_1 = \vec{d}_2$. In this case, the transition dipole moment operators are expressed in the exciton basis as $\mu^+ = \sqrt{2}(\hat{\epsilon}\cdot\vec{d})\ket{E_b}\bra{g}$ and $\mu^- = \sqrt{2}(\hat{\epsilon}\cdot\vec{d})\ket{g}\bra{E_b}$. This implies that for a positive-valued electronic coupling $V$, a high-energy exciton state $\ket{E_b}=\frac{1}{\sqrt{2}}(\ket{e_1}+\ket{e_2})$ is bright, whereas a low-energy exciton state $\ket{E_d}=\frac{1}{\sqrt{2}}(-\ket{e_1}+\ket{e_2})$ is dark as it is not directly excited by an optical pulse described by $\hat{\mu}^+$, namely $\bra{E_d}\mu^+\ket{g}=0$. The lifetime of the optical coherence $\ket{E_b}\bra{g}$ determines the width of an absorption line shape, which is governed by the interaction between electronic states and vibrational modes.

\subsection{Second order cumulant expansion}

\begin{figure*}[t]
	\includegraphics[width=1\textwidth]{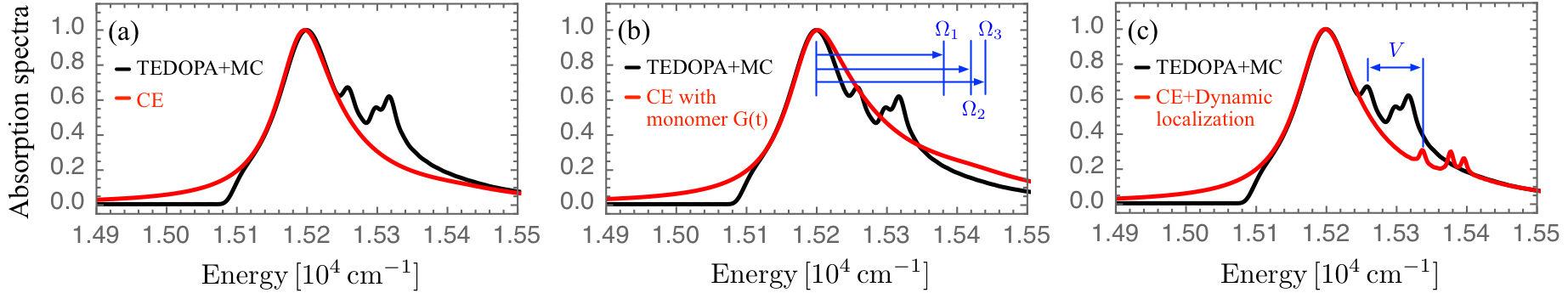}
	\caption{(a) Numerically exact absorption spectrum computed by TEDOPA+MC is shown in black, and approximate results obtained by the standard line shape theory based on second order cumulant expansion (CE) are shown in red. (b) Approximate absorption spectrum computed by CE is shown in red, where a prefactor $\sum_{i=1}^{2}|\langle e_i|E_{j}\rangle|^{4}=1/2$ suppressing the intensity of a vibrational sideband is ignored. (c) Approximate results obtained by a modified line shape theory with dynamic localization are shown in red. We note that the maximum amplitude of every absorption spectrum is normalized to unity for a comparison of different models.}
	\label{FigS1_JL}
\end{figure*}

In the line shape theory based on second order cumulant expansion~\cite{Mukamel1995,NovoderezhkinJPCB2004,renger2006,AbramaviciusJCP2010}, the absorption spectrum is approximately described by
\begin{align}
	A(\omega)\propto {\rm Re}\left[\omega\sum_{j=b,d}|\bra{E_j}\mu^+\ket{g}|^{2}\int_{0}^{\infty}dt \exp\left(i(\omega-E_{j}-E_{{\rm LS},j})t+G_{j}(t)-G_{j}(0)-\left(\frac{1}{2}\gamma_{{\rm rxn},j}+\gamma_{\rm pd}\right)t\right)\right],
\end{align}
where $|\bra{E_j}\mu^+\ket{g}|^{2}$ is the transition dipole strength of an exciton state $\ket{E_j}$, which is zero for the dark low-energy exciton $\ket{E_d}$ in our model. Moreover, $E_j$ and $E_{{\rm LS},j}$ represent the exciton energy of $\ket{E_j}$ and its shift induced by vibronic couplings, described by a second order perturbation theory~\cite{breuer02,Felipe2021}
\begin{equation}
	E_{{\rm LS},j}=-\sum_{i=1}^{2}|\langle e_i|E_{j}\rangle|^{4}\lambda+\frac{1}{\pi}\sum_{i=1}^{2}|\langle E_{b}|e_i\rangle\langle e_i|E_{d}\rangle|^{2}\,{\rm P}\int_{-\infty}^{\infty}d\omega\,\frac{J(\omega)(n(\omega)+1)+J(-\omega)n(-\omega)}{\Delta E_{j}-\omega},
\end{equation}
where $\lambda=\int_{0}^{\infty}d\omega J(\omega)/(\pi\omega)$ is the reorganization energy, $\Delta E_{j}=E_{j}-E_{k}$ with $j\neq k$, representing an excitonic splitting, and ${\rm P}$ denotes the Cauchy principal value; $\gamma_{{\rm rxn},j}$ is the relaxation rate from $\ket{E_{0,j}}$ to $\ket{E_{0,k}}$ with $j\neq k$, described by a second order perturbation theory~\cite{breuer02,Felipe2021}
\begin{equation}
	\gamma_{{\rm rxn},j}=2 \sum_{i=1}^{2}|\langle E_{j}|e_i\rangle\langle e_i|E_{k}\rangle|^{2}(J(\Delta E_{j})(n(\Delta E_{j})+1)+J(-\Delta E_{j})n(-\Delta E_{j})),
\end{equation}
and $\gamma_{\rm pd}$ is a phenomenological pure dephasing rate that is taken to be zero unless stated otherwise. A vibrational sideband in the line shape theory originates from $G_{j}(t)=\sum_{i=1}^{2}|\langle e_i|E_{j}\rangle|^{4} G(t)$ with $G(t)$ defined by
\begin{equation}
	G(t)=\frac{1}{\pi}\int_{0}^{\infty}d\omega \,\omega^{-2}\left(J(\omega)(n(\omega)+1)e^{-i\omega t}+J(\omega)n(\omega)e^{i\omega t}\right),
\end{equation}
where $n(\omega)=(\exp(\omega/k_B T)-1)^{-1}$. For a monomer, optical coherence dynamics can be computed exactly by using an analytical solution: ${\rm Tr}[\mu^- \mathcal{U}(t)[\mu^+ |g\rangle\langle g|\otimes\rho_{v}(T)]\propto \exp(-i (E-\lambda) t+G(t)-G(0))$. In the case of a dimer, the line shape theory only provides an approximate solution of optical coherence dynamics, and it is notable that the vibrational sideband is determined by $G_{j}(t)-G_{j}(0)=\sum_{i=1}^{2}|\langle e_i|E_{j}\rangle|^{4} (G(t)-G(0))$ whose amplitude is suppressed by the delocalization of the exciton states over two sites, $\sum_{i=1}^{2}|\langle e_i|E_{j}\rangle|^{4}=1/2$. In Fig.~\ref{FigS1_JL}(a), a numerically exact absorption spectrum computed by TEDOPA+MC, shown in black, is compared with an approximate absorption line shape obtained by the second order cumulant expansion, shown in red, where a vibrational sideband observed in TEDOPA+MC simulations is almost not visible in the approximate results. The absence of the vibrational sideband in the approximate results is partly due to the fact that the intensity of the vibrational sideband is suppressed by the exciton delocalization in the line shape theory, namely a prefactor $\sum_{i=1}^{2}|\langle e_i|E_{j}\rangle|^{4}=1/2$ in $G_{j}(t)-G_{j}(0)=\sum_{i=1}^{2}|\langle e_i|E_{j}\rangle|^{4} (G(t)-G(0))$. As shown in Fig.~\ref{FigS1_JL}(b), even if the prefactor is ignored in the line shape theory, the vibrational sideband in numerically exact results around $\sim 15300\,{\rm cm}^{-1}$, shown in black, cannot be reproduced by the approximate results, shown in red, as the vibrational sideband of the bright high-energy exciton state $\ket{E_b}$ appears in a region around $\sim 15400\,{\rm cm}^{-1}$. In the line shape theory, the energy-gap between main absorption peak centered at $\sim 15200\,{\rm cm}^{-1}$ and vibrational sideband is determined by the vibrational frequencies of the intrapigment modes, $(\Omega_1,\Omega_2,\Omega_3)=(181,221,240)\,{\rm cm}^{-1}$, as highlighted by blue arrows in Fig.~\ref{FigS1_JL}(b). The vibrational sideband in the standard line shape theory is a broad peak, contrary to the three narrow peaks in numerically exact results, which is due to the fact that their widths are determined by the relaxation rate $\gamma_{{\rm rxn},b}\approx 91\,{\rm cm}^{-1}$ from the bright high-energy exciton $\ket{E_b}$ to the dark low-energy exciton state $\ket{E_d}$.

A modified line shape theory based on the idea of dynamic localization~\cite{RengerJPP2011,MuhBA2012} also cannot reproduce the numerically exact absorption spectrum, as shown in Fig.~\ref{FigS1_JL}(c). The peak positions of the vibrational sideband in approximate results, shown in red, are blue-shifted from those in numerically exact results, shown in black. This is due to the fact that in the modified line shape theory, the vibrational sideband of a dimer is replaced by the vibrational sideband of a monomer~\cite{Felipe2021} and as a result the peak positions of the vibrational sideband are approximately given by $E+\Omega_k$ where $E$ is the energy-level of a monomer and $\Omega_k$ are the vibrational frequencies of the three intrapigment modes. Contrary to the dynamic localization theory, the vibrational sideband in the numerically exact absorption spectrum originates from vibrationally hot dark exciton states $\ket{E_d,1_k}$ borrowing transition dipole strengths from a vibrationally cold bright exciton state $\ket{E_b,0}$ where $\ket{0}$ represents a global vibrational ground state where all the intrapigment modes are in their vacuum states, and $\ket{1_k}$ is a singly-excited vibrational state where only the $k$-th intrapigment mode is singly excited and the other two modes are in their vacuum states (see Section \ref{section_Markov_JL} for more detail). As a result the peak positions of the vibrational sideband in numerically exact results are approximately given by $E_d+\Omega_k=E-V+\Omega_k$ where $E_d$ is the energy-level of the dark exciton state $\ket{E_d}$ and $V=69\,{\rm cm}^{-1}$ is the electronic coupling considered in simulations. This is the reason why the vibrational sideband in the line shape theory with dynamic localization ($E+\Omega_k$) is blue-shifted by $\sim 80\,{\rm cm}^{-1}$ from that of the numerically exact results obtained by TEDOPA+MC ($E_d+\Omega_k=E-V+\Omega_k$), as highlighted by a blue arrow in Fig.~\ref{FigS1_JL}(c).

\subsection{Reduced vibronic model with Markovian description of a broad phonon spectrum}\label{section_Markov_JL}

We now consider a reduced vibronic model where three intrapigment vibrational modes are included in system Hamiltonian in addition to the electronic states~\cite{LimNC2015,Felipe2021}. The remaining broad phonon spectrum $J_{\rm AR}(\omega)$ is approximately treated by using a Lindblad equation, inducing Markovian noise~\cite{Felipe2021}. The vibrational damping of the intrapigment modes is also included in the construction of the Lindblad equation by considering a secondary bath coupled to each intrapigment mode~\cite{Felipe2021}. The Lindblad equation is constructed in the vibronic eigenbasis of the system Hamiltonian where the pure dephasing and relaxation rates of the vibronic eigenstates are determined by the broad phonon spectrum $J_{\rm AR}(\omega)$ and an effective coupling spectrum of the secondary bath, as detailed below. We consider only the modes $b_k$ that describe relative motion in the Lindblad equation, as the contribution of the center-of-mass motion modes $B_k$ to absorption and 2D electronic spectra can be treated in an analytical way~\cite{Felipe2021}.

In this reduced model, the total Hamiltonian is decomposed into four parts, $H=H_0+H_1+H_2+H_c$ where $H_c$ is the Hamiltonian of the center-of-mass modes in Eq.~(\ref{eq:lim_H_c}). The system Hamiltonian $H_0$ consists of the electronic Hamiltonian $H_e$ and the vibrational Hamiltonian of the three intrapigment modes with vibrational frequencies $\Omega_k$ and Huang-Rhys factors $S_k$
\begin{equation}
	H_0=H_e+\sum_{k=1}^{3}\Omega_k b_k'^\dagger b_k'-(\ket{E_b}\bra{E_d}+\ket{E_d}\bra{E_b})\sum_{k=1}^{3}\Omega_{k}\sqrt{\frac{S_k}{2}}(b_{k}'^{\dagger}+b_{k}'),
\end{equation}
where $b_{k}'$ describes the relative motion of the $k$-th intrapigment modes. In simulations, we consider a composite vibrational basis $\{\ket{n_1,n_2,n_3}\}$ where $n_k$ denotes the number of vibrational excitations in the mode $b_{k}'$. To check the convergence of numerical data, such as computed absorption spectra, we consider  $N$ vibrational excitation subspace spanned by the composite vibrational states with $0\le \sum_{k=1}^{3}n_k\le N$ and increase the value of $N$ until the data show numerical convergence. To compute the dynamics of the vibronic system, we consider a Lindblad equation where the pure dephasing and relaxation rates of the vibronic eigenstates of the system Hamiltonian $H_0$ are determined by $H_1+H_2$ in the total Hamiltonian~\cite{breuer02,Felipe2021}
\begin{align}
	H_1 &= \sum_{k}\omega_{k}b_{k}''^{\dagger}b_{k}''-(\ket{E_b}\bra{E_d}+\ket{E_d}\bra{E_b})\sum_{k}\omega_{k}\sqrt{\frac{s_k}{2}}(b_{k}''^{\dagger}+b_{k}''),\\
	H_2 &= \sum_{k=1}^{3}\sum_{l}\omega_{l}c_{k,l}^{\dagger}c_{k,l}+\sum_{k=1}^{3}(b_{k}'^\dagger+b_{k}')\sum_{l}g_{k,l}(c_{k,l}^\dagger+c_{k,l}).
\end{align}
Here, $H_1$ is the Hamiltonian of the relative motion modes $b_{k}''$ of the broad phonon spectrum $J_{\rm AR}(\omega)$, which is not included in the vibronic system Hamiltonian $H_0$, while $H_2$ describes a secondary bath coupled to each intrapigment mode, modelled by $b_{k}'$ in $H_0$, where vibrational excitations are exchanged between intrapigment modes $b_{k}'$ and secondary bath modes $c_{k,l}$. The coupling spectrum of the $k$-th secondary bath, characterized by $g_{k,l}$, is modelled by a Lorentzian spectral density centered at the vibrational frequency $\Omega_k$ of the $k$-th intrapigment mode, so that the coupling strength is maximized when environmental mode frequencies $\omega_l$ are near-resonant with $\Omega_k$ of the intrapigment mode. The width of the Lorentzian function is taken to be $(50\,{\rm fs})^{-1}$, so that the corresponding secondary-bath correlation function quickly decays within the time scale of the vibrational damping of the intrapigment modes, $\sim(500\,{\rm fs})^{-1}$. The height of the Lorentzian function is determined in such a way that the Lindblad damping rate of the $k$-th mode is identical to the vibrational damping rate, $\sim(500\,{\rm fs})^{-1}$, considered in TEDOPA+MC simulations, when the $k$-th mode is decoupled from electronic states, namely when the single vibrational mode is coupled to only its secondary bath. See Ref.~\cite{Felipe2021} for more detail.

In absorption simulations, one needs to compute the dipole-dipole correlation function, ${\rm Tr}[\mu^- \mathcal{U}(t)[ \mu^+ |g\rangle\langle g|\otimes \rho_v(T)]]$. The thermal state of the total vibrational environments can be decomposed into four parts, $\rho_v(T)=\rho_{v,r}'(T)\otimes \rho_{v,r}''(T) \otimes \rho_{v,c}(T) \otimes \rho_{s}(T)$ where $\rho_{v,r}'(T)$ and $\rho_{v,r}''(T)$ are the thermal states of the relative motions of the three intrapigment modes and the broad phonon spectrum $J_{\rm AR}(\omega)$, respectively, and $\rho_{v,c}(T)$ is the thermal state of the center-of-mass motion modes of the total spectral density $J(\omega)$. $\rho_{s}(T)$ is the thermal state of the secondary baths. In reduced model simulations, $\rho_{v,r}'(T)$ is included in a reduced system density matrix, while $\rho_{v,r}''(T)$ and $\rho_{s}(T)$ are considered in the construction of the Lindblad equation (Markovian noise). It can be shown that the state of the center-of-mass modes is separable from the other degrees of freedom, namely electronic states and relative motion modes, as the center-of-mass modes are coupled to the electronic states via $(\sum_{i=1}^{2}\ket{e_i}\bra{e_i})\otimes \sum_{k}\omega_k \sqrt{s_k/2}(B_k+B_k^\dagger)$ where $\sum_{i=1}^{2}\ket{e_i}\bra{e_i}$ is an identity operator within single electronic excitation subspace. As a result, the center-of-mass modes can dephase optical coherence between electronic ground and excited states, but do not affect electronic dynamics within the single electronic excitation subspace. The contribution of the center-of-mass modes to absorption spectrum can be taken into account exactly by using the analytical solution of optical coherence dynamics of a monomer coupled to a vibrational environment with an effective spectral density $J(\omega)/2$~\cite{Felipe2021}.

\begin{figure*}[t]
	\includegraphics[width=1\textwidth]{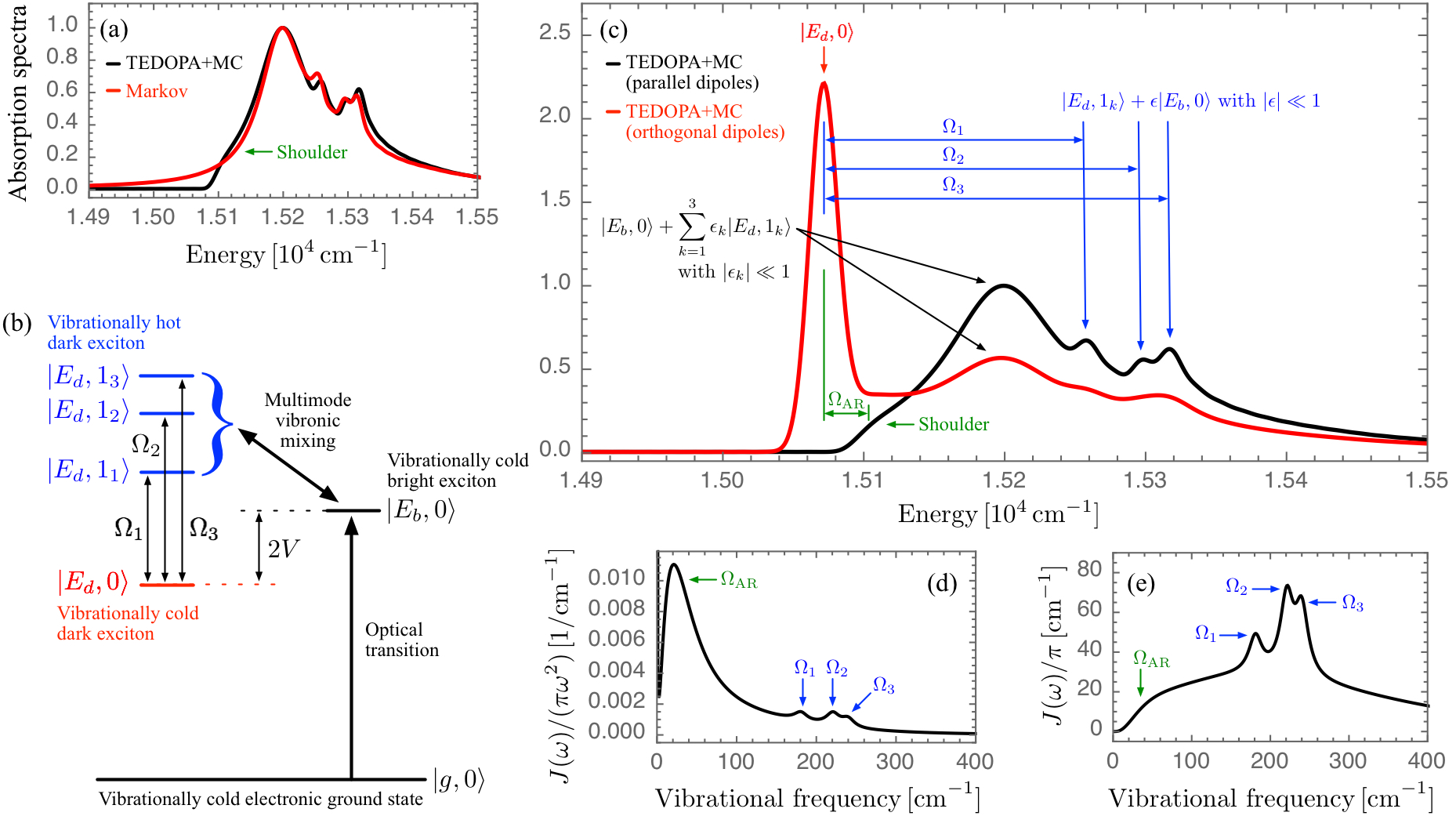}
	\caption{(a) Numerically exact absorption spectrum computed by TEDOPA+MC, shown in black, and approximate absorption line shape obtained by a reduced model, shown in red, where a broad phonon spectrum is modelled by a Lindblad equation (Markovian noise). (b) Energy-level structure of a vibronic system Hamiltonian where three intrapigment modes are included in system Hamiltonian in addition to electronic states. (c) Numerically exact absorption spectrum of a dimer where monomer transition dipoles $\vec{d}_1$ and $\vec{d}_2$ are parallel (orthogonal) to each other, shown in black (red). (d) Huang-Rhys factor distribution $J(\omega)/(\pi\omega^2)$. (e) Spectral density $J(\omega)$. We note that in (a), the maximum amplitudes of numerically exact and approximate absorption spectra are independently normalized and taken to be unity for a comparison of different models. In (c), the maximum amplitudes of the parallel and orthogonal cases are normalized by a common factor to demonstrate that the absorption peak at $\sim 15200\,{\rm cm}^{-1}$, originating from a vibrationally cold high-energy exciton state $\ket{E_b,0}$, has a smaller amplitude for the orthogonal case than for the parallel case. In the orthogonal case, the total transition dipole strength of two monomers is distributed to two exciton states, while in the parallel case, the high-energy exciton state carries all the transition dipole strength as the low-energy exciton state becomes dark. This is the reason why the amplitude of the absorption peak at $\sim 15200\,{\rm cm}^{-1}$ becomes smaller for the orthogonal case than for the parallel case. A Gaussian dephasing considered in the orthogonal case also reduces the absorption peak intensity, but this effect is minor than the dipole distribution effect.}
	\label{FigS2_JL}
\end{figure*}

In Fig.~\ref{FigS2_JL}(a), the absorption spectrum computed by the reduced vibronic model is shown in red, where the peak positions of the vibrational sideband are well matched to numerically exact results shown in black. It is found that the reduced model results obtained for double ($N=2$) and triple ($N=3$) vibrational excitation subspaces are well matched (not shown here). By analyzing the vibronic eigenstates of the system Hamiltonian $H_0$, it is found that the main absorption peak originates from a vibrationally cold bright exciton state $\ket{E_b,0}$ that is vibronically mixed with vibrationally hot dark exciton states $\ket{E_d,1_k}$ with $k\in\{1,2,3\}$, mediated by the off-diagonal vibronic couplings, as schematically shown in Fig.~\ref{FigS2_JL}(b). The three peaks in the vibrational sideband originate from the vibrationally hot dark exciton states $\ket{E_d,1_k}$ that barrow transition dipole strengths from the bright exciton state $\ket{E_b,0}$, mediated by a vibronic mixing, and become visible in absorption spectrum.

This interpretation is further supported by additional TEDOPA+MC simulations where transition dipole moments $\vec{d}_i$ are taken to be mutually orthogonal to each other. In this case, the transition dipole strengths of both exciton states, $|\bra{E_b}\mu^+\ket{g}|\propto |\hat{e}\cdot(\vec{d}_1+\vec{d}_2)|$ and $|\bra{E_d}\mu^+\ket{g}|\propto |\hat{e}\cdot(\vec{d}_1-\vec{d}_2)|$, become non-zero when an orientational average of the dipole moments $\vec{d}_i$ with respect to the polarisation $\hat{e}$ of an external field is considered in simulations (isotropic ensemble), meaning that both excitons are bright and can be directly excited by an optical pulse. At zero temperature, the lifetime of the optical coherence related to the low-energy exciton state, $\ket{E_d}\bra{g}$, is extremely long due to a slow relaxation from low- to high-energy exciton states, and the coherence does not decay close to zero within a simulation time of $\sim 2\,{\rm ps}$. In order to estimate the absorption peak position related to a vibrationally cold low-energy exciton state $\ket{E_d,0}$, we multiply a Gaussian dephasing $\exp(-\Delta^2 t^2)$ with $\Delta=7\,{\rm cm}^{-1}$ to the optical coherence dynamics computed by TEDOPA+MC, which corresponds to a correlated static disorder where the site energies $E_i$ of two monomers are shifted in a correlated way for each sub-ensemble~\cite{Felipe2021}. The additional Gaussian dephasing makes the resultant optical coherence dynamics decay to zero within $\sim 2\,{\rm ps}$, leading to a well-defined absorption line shape when Fourier transformed. As highlighted by a red arrow in Fig.~\ref{FigS2_JL}(c), a narrow low-energy absorption peak appears when the monomer dipole moments $\vec{d}_1$ and $\vec{d}_2$ are orthogonal to each other, as shown in red, and this low-energy peak is absent when the dipole moments are parallel, as shown in black. This is due to the fact that the low-energy exciton is bright and dark, respectively, in the orthogonal and parallel cases. It is found that the energy-gaps between the narrow lowest-energy absorption peak in the orthogonal case and the three narrow peaks in a vibrational sideband, observed in both orthogonal and parallel cases, are close to the vibrational frequencies $\Omega_k$ of the three intrapigment modes, as highlighted by blue arrows in Fig.~\ref{FigS2_JL}(c). This implies that these absorption peaks originate from vibrationally cold and hot low-energy exciton states, $\ket{E_d,0}$ and $\ket{E_d,1_k}$ with an energy-gap of $\Omega_k$, although the peak positions of the vibrational sideband can be slightly shifted by a vibronic mixing of $\ket{E_d,1_k}$ with $\ket{E_b,0}$~\cite{LimNC2015}.

However, the low-energy part of absorption spectrum cannot be reproduced by the reduced model. As highlighted by a green arrow in Fig.~\ref{FigS2_JL}(a) and (c), there is a shoulder in the low-energy part of numerically exact absorption line shape. The shoulder observed in the parallel dipole case is blue-shifted by $\sim 30\,{\rm cm}^{-1}$ from the lowest-energy absorption peak observed in the orthogonal dipole case, as highlighted by a green arrow in Fig.~\ref{FigS2_JL}(c). This suggests the possibility that the dark exciton state in the presence of vibrational excitations of the broad phonon spectrum $J_{{\rm AR}}(\omega)$ could be responsible for the low-energy shoulder. Fig.~\ref{FigS2_JL}(d) shows the distribution of the Huang-Rhys factors of the total spectral density, $J(\omega)/(\pi\omega^2)=\sum_{k}s_k \delta(\omega-\omega_k)$, which determines the transition dipole strength of a vibrational sideband in the case of a monomer~\cite{Felipe2021}. It is notable that the broad phonon spectrum $J_{\rm AR}(\omega)$ leads to a relatively narrow Huang-Rhys factor distribution $J_{\rm AR}(\omega)/(\pi\omega^2)$ with a peak around $\sim 30\,{\rm cm}^{-1}$ (see the spectral density $J(\omega)$ shown in Fig.~\ref{FigS2_JL}(e) where $J_{\rm AR}(\omega)$ leads to a broad phonon spectrum).

\subsection{Reduced vibronic model with non-Markovian description of a broad phonon spectrum}\label{section_nonMarkov_JL}

\begin{figure*}[t]
	\includegraphics[width=1\textwidth]{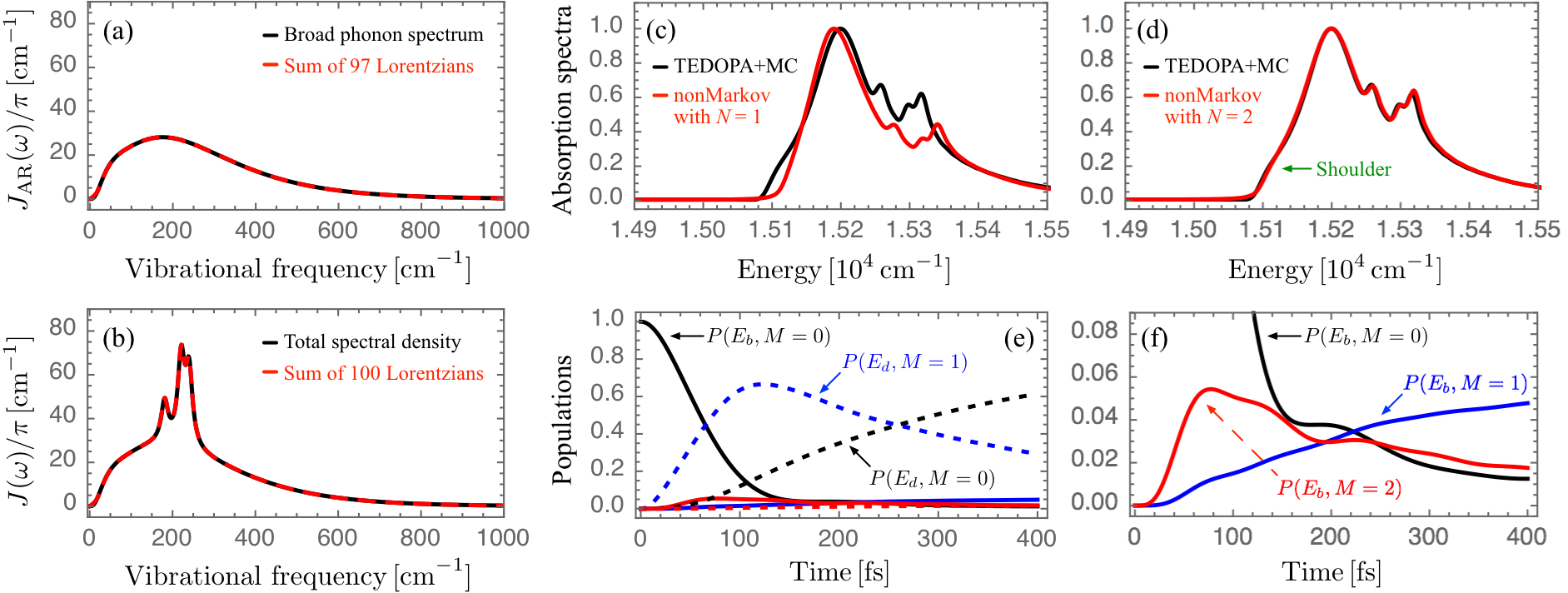}
	\caption{(a) Broad phonon spectral density $J_{\rm AR}(\omega)$, shown in black, and a sum of 97 Lorentzian functions, shown in red. (b) Total spectral density $J(\omega)$, shown in black, and a sum of 100 Lorentzian functions, shown in red, including the 97 Lorentzian functions considered in (a). (c) Numerically exact absorption spectrum computed by TEDOPA+MC, shown in black, and approximate absorption spectrum computed by a reduced model, shown in red, where 100 effective damped quantum harmonic oscillators are considered within single $(N=1)$ vibrational excitation subspace. (d) Approximate results obtained by the reduced model, shown in red, where double $(N=2)$ vibrational excitation subspace is considered. (e) Time evolution of the population $P(E_i,M=0)$ of a vibrationally cold $i$-th exciton state $\ket{E_i,0}$, shown in black, and the total populations $P(E_i,M=1)$ and $P(E_i,M=2)$ of singly- ($\ket{E_i,1_j}$) and doubly-excited vibrational states ($\ket{E_i,1_j,1_k}$ and $\ket{E_i,2_j}$) of the $i$-th exciton state, shown in blue and red, respectively. The populations of the bright ($i=b$) and dark ($i=d$) exciton states are shown in solid and dashed lines, respectively. (f) The population dynamics of the bright singly- and doubly-excited vibrational states are shown in detail.}
	\label{FigS3_JL}
\end{figure*}

To investigate the influence of vibrational excitations of the broad phonon spectrum $J_{\rm AR}(\omega)$ on absorption line shapes, we consider a sum of 100 Lorentzian spectral densities that is well fitted with the total spectral density $J(\omega)$. As displayed in Fig.~\ref{FigS3_JL}(a), the broad phonon spectrum $J_{\rm AR}(\omega)$, shown in black, is fitted with a sum of 97 Lorentzian functions, shown in red. The total spectral density $J(\omega)$, including three intrapigment modes modelled by three narrow Lorentzian functions, is therefore fitted with a sum of 100 Lorentzian functions, as shown in Fig.~\ref{FigS3_JL}(b). For simplicity, we consider a uniform width of the 100 Lorentzian functions and take the value of the vibrational damping rate, $\sim 500\,{\rm fs}$, of the three intrapigment modes considered in TEDOPA+MC simulations. In reduced model simulations, we consider each Lorentzian function as a quantum harmonic oscillator under a local Lindblad damping
\begin{equation}
	\frac{d}{dt}\rho(t)=-i[H_{\rm non-Markov},\rho(t)]+\sum_{k=1}^{100}\gamma \left(2 a_k \rho(t) a_k^\dagger-\{a_k^\dagger a_k,\rho(t)\}\right),
\end{equation}
with $\gamma\approx (500\,{\rm fs})^{-1}$, $\hat{a}_k^{\dagger}$ and $\hat{a}_k$ represent the creation and annihilation operators, respectively, of the $k$-th effective mode. The vibrational frequency, Huang-Rhys factor and damping rate of each effective mode are determined, respectively, by the peak position, height and width of the corresponding Lorentzian function. The reduced model Hamiltonian $H_{\rm non-Markov}$ is constructed in a composite vibrational basis $\{\ket{n_1,n_2,n_3,\cdots,n_{100}}\}$ with $n_k$ denoting the number of vibrational excitations of the $k$-th effective mode. We note that these effective modes describe the relative motion modes of the spectral density $J(\omega)$, as the contribution of the center-of-mass modes to optical line shapes can be taken into account analytically. We consider $N$ vibrational excitation subspace spanned by the composite vibrational states with $0\le \sum_{k=1}^{100} n_k\le N$. Fig.~\ref{FigS3_JL}(c) and (d) show absorption spectra computed by the reduced model within single ($N=1$) and double ($N=2$) vibrational excitation subspaces, respectively. It is notable that the doubly-excited vibrational states are essential to reproduce numerically exact absorption line shape obtained by TEDOPA+MC, and the low-energy shoulder in absorption spectrum can be well reproduced by the reduced model. These results imply that the initial bright exciton state $\ket{E_b,0}$, where all the vibrational modes are in their ground states, is mixed with vibrationally hot dark states $\ket{E_d,1_j}$, and then these singly-excited vibrational states $\ket{E_d,1_j}$ are subsequently mixed with the doubly-excited vibrational states $\ket{E_b,1_j,1_k}$ and $\ket{E_b,2_j}$ where different $j$-th and $k$-th modes are singly excited, or the $j$-th mode is doubly excited. As shown in Fig.~\ref{FigS3_JL}(e), the population $P(E_b,M=0)$ of the initial bright exciton state $\ket{E_b,0}$, shown in a black solid line, decays on a $100\,{\rm fs}$ time scale, leading to an increase in the total population $P(E_d,M=1)$ of the dark singly-excited vibrational states $\ket{E_d,1_j}$, as shown in a blue dashed line, and then the damping of vibrational modes generates the population $P(E_d,M=0)$ of a vibrationally cold dark exciton state $\ket{E_d,0}$, as shown in a black dashed line, with $M$ denoting the total number of vibrational excitations of the 100 effective modes. As shown in Fig.~\ref{FigS3_JL}(f), the dark singly-excited vibrational states $\ket{E_d,1_j}$ are coupled to the bright doubly-excited vibrational states, $\ket{E_b,1_j,1_k}$ and $\ket{E_b,2_j}$, and increase the total population $P(E_b,M=2)$ of the doubly-excited states, as shown in a red solid line. The vibrational damping process decreases the population $P(E_b,M=2)$ of the doubly-excited states and enhances the population $P(E_b,M=1)$ of the bright singly-excited vibrational states $\ket{E_b,1_j}$, as shown in a blue solid line. It is notable that the total population $P(E_b,M=2)$ of the doubly-excited vibrational states is not negligible, as it is of the order of $\sim 5\,\%$. Importantly, the population $P(E_b,M=0)$ of the initial bright state $\ket{E_b,0}$ becomes comparable to the total population $P(E_b,M=2)$ of the bright doubly-excited vibrational states $\ket{E_b,1_j,1_k}$ and $\ket{E_b,2_j}$ within a 100\,fs time scale, $t\ge 150\,{\rm fs}$. This implies that when the electronic-vibrational dynamics is monitored by a pump-probe scheme, the bright doubly-excited vibrational states $\ket{E_b,1_j,1_k}$ and $\ket{E_b,2_j}$ can dominate nonlinear optical spectra when the time delay between pump and probe is sufficiently long enough to suppress the population of the vibrationally cold initial bright state $\ket{E_b,0}$. The high population of the singly-excited vibrational states $\ket{E_d,1_j}$ cannot be directly probed by an optical pulse as they are dark states.

\section{Two-dimensional electronic spectroscopy} \label{sec:2d}

To investigate the contribution of the doubly-excited vibrational states $\ket{E_b,1_j,1_k}$ and $\ket{E_b,2_j}$ to nonlinear optical responses involving long-time vibronic dynamics, we consider 2D electronic spectroscopy~\cite{JonasARPC2003,BrixnerJCP2004}. In this modified pump-probe scheme, a molecular sample is perturbed by two pump pulses and one probe pulse with controlled time delays. When the wave vectors of the three optical pulses are denoted by $\vec{k}_1$, $\vec{k}_2$, $\vec{k}_3$, respectively, nonlinear optical signals can be measured at several phase-matched directions $\pm\vec{k}_1 \pm\vec{k}_2 \pm\vec{k}_3$. In this work, we consider rephasing 2D spectra measured at a phase-matched direction $-\vec{k}_1+\vec{k}_2+\vec{k}_3$. In 2D experiments, the intensity of laser pulses is reduced until measured nonlinear signals show convergence, so that the measured 2D spectra are dominated by a third-order optical response with relatively weaker contributions of higher-order nonlinear optical signals. Such a weak light-matter interaction can be well described by a perturbation theory and it can be shown that the rephasing 2D signals originate from three distinct pulse-driven molecular dynamics, called ground state bleaching (GSB), stimulated emission (SE) and excited state absorption (ESA)~\cite{Mukamel1995,LimPRL2019}. The GSB and SE signals originate from optical coherence dynamics between electronic ground and singly-excited states (excitons), while the ESA signal involves the optical coherence between singly- and doubly-excited electronic states (biexcitons). Contrary to the GSB signal, where vibrational dynamics in the electronic ground state manifold is monitored as a function of the time delay $t_2$ between second pump and probe pulses, the SE and ESA signal provide the information about electronic-vibrational (vibronic) dynamics in the electronic excited state manifold as a function of the waiting time $t_2$. In this work, we consider the SE component of the rephasing spectra where two-dimensional optical line shapes in the excitation $\omega_1$ and detection $\omega_3$ energy domain evolve as a function of the waiting time $t_2$ due to the vibronic dynamics in the electronic excited state manifold. For simplicity, we consider a broad bandwidth of the pump and probe pulses, and compute the SE component in the impulsive limit where the temporal profiles of the laser pulses are approximately described by a Dirac delta function in the time domain~\cite{Mukamel1995,LimPRL2019}. The SE component of the rephasing signal, $R_{{\rm SE}}(\omega_1,t_2,\omega_3)$, is described by
\begin{equation}
	R_{\rm SE}(\omega_1,t_2,\omega_3) \propto \int_{0}^{\infty}dt_1 e^{-i\omega_1 t_1}\int_{0}^{\infty} dt_3 e^{i\omega_3 t_3} {\rm Tr}[\mu^- \mathcal{U}(t_3)[\mathcal{U}(t_2)[\mu^+ \mathcal{U}(t_1)[\ket{g}\bra{g}\otimes\rho_{v}(T)\mu^-]]\mu^+]].\label{eq:SE_JL}
\end{equation}
Here, $\ket{g}\bra{g}\otimes\rho_{v}(T)\mu_{{\rm pump},1}^-$ describes a molecular sample perturbed by the first pump pulse, creating an optical coherence between electronic ground and excited states. The coherence evolves during a time-delay $t_1$ between first and second pump pulses, represented by a time evolution operator $\mathcal{U}(t_1)[\sigma]=e^{-iHt_1}\sigma e^{iHt_1}$ with the total Hamiltonian $H$. The second pump pulse induces an optical transition from electronic ground to excited states in the ket space, $\mu_{{\rm pump},2}^+ \mathcal{U}(t_1)[\ket{g}\bra{g}\otimes\rho_{v}(T)\mu_{{\rm pump},1}^-]$, which enables one to monitor vibronic dynamics in the electronic excited state manifold as a function of the waiting time $t_2$ between second pump and probe pulses, represented by $\mathcal{U}(t_2)[\sigma]=e^{-iHt_2}\sigma e^{iHt_2}$. The probe pulse induces an optical transition from electronic excited to ground states in the bra space, $\mathcal{U}(t_2)[\mu_{{\rm pump},2}^+ \mathcal{U}(t_1)[\ket{g}\bra{g}\otimes\rho_{v}(T)\mu_{{\rm pump},1}^-]]\mu_{{\rm probe}}^+$, leading to an oscillatory average dipole moment as a function of time $t_3$, which generates a third order optical response depending on the controlled time delays $t_1$ and $t_2$ amongst pulses. In simulations, a two-dimensional optical line shape is obtained by Fourier transforming the third order optical response with respect to $t_1$ and $t_3$, leading to excitation $\omega_1$ and detection $\omega_3$ energies, respectively.

\begin{figure*}[t]
	\includegraphics[width=1\textwidth]{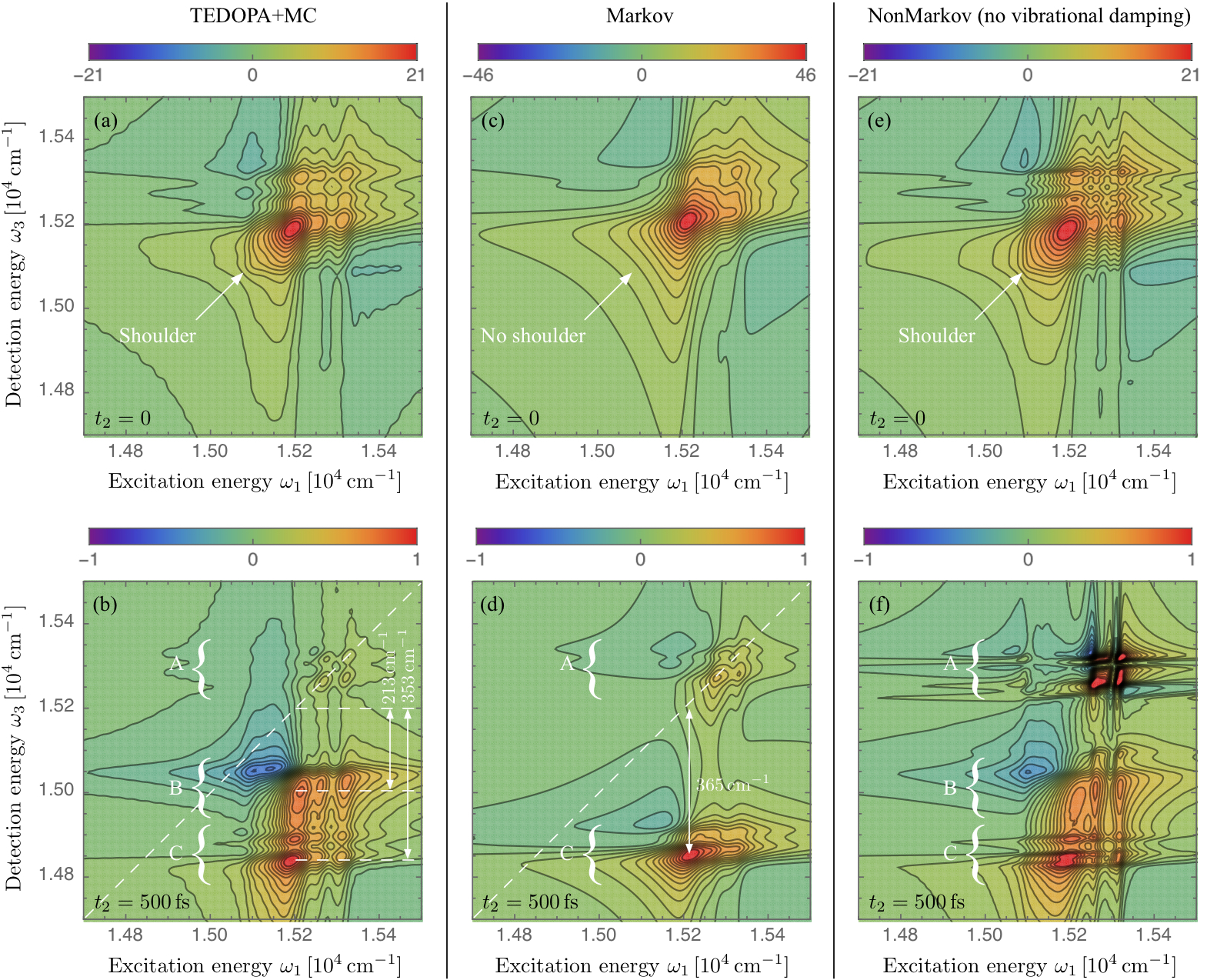}
	\caption{(a,b) The real part of the SE component of rephasing 2D spectra at $t_2=0$ and $t_2=500\,{\rm fs}$ computed by TEDOPA+MC. (c,d) 2D line shapes at $t_2=0$ and $t_2=500\,{\rm fs}$ computed by a reduced model where a broad phonon spectrum $J_{\rm AR}(\omega)$ is approximately described by a Lindblad equation (Markovian noise). (e,f) 2D line shapes at $t_2=0$ and $t_2=500\,{\rm fs}$ computed by a reduced model where the total spectral density $J(\omega)$ is approximately described by 100 undamped vibrational modes (non-Markovian noise). For simplicity, exponential dephasing for $t_1$ and $t_3$ dynamics is considered instead of taking into account the vibrational damping of the 100 effective modes in simulations.}
	\label{FigS4_JL}
\end{figure*}

Fig.~\ref{FigS4_JL}(a) and (b) show the real part of the SE component of the rephasing spectra at $t_2=0$ and $t_2=500\,{\rm fs}$, respectively, computed by TEDOPA+MC. The positions of 2D peaks at $t_2=0$ are well matched to the absorption peak positions computed by TEDOPA+MC shown in Fig.~\ref{FigS3_JL}(d). It is notable that the overall intensity of 2D spectra at $t_2=0$ is an order of magnitude stronger than that of 2D data at $t_2=500\,{\rm fs}$. This is due to the population transfer from initial bright exciton state $\ket{E_b,0}$ to dark states $\ket{E_d,1_j}$ (see Fig.~\ref{FigS3_JL}(e)) where a high population of the optically dark states results in a suppressed 2D signal intensity at $t_2=500\,{\rm fs}$. The 2D peaks at $t_2=500\,{\rm fs}$ can be categorized into three regions A, B, C, as highlighted in Fig.~\ref{FigS4_JL}(b). Importantly, the 2D peaks in the regions B and C are below diagonal $(\omega_1>\omega_3)$ and the differences in excitation and detection energies, $|\omega_1-\omega_3|$, of these cross peaks are in the range of $200\sim 400\,{\rm cm}^{-1}$, as marked by white arrows in Fig.~\ref{FigS4_JL}(b), which are two times larger than the vibrational frequencies of the spectral density $J(\omega)$, shown in Fig.~\ref{FigS2_JL}(d) and (e). These results suggest the possibility that the cross peaks in the regions B and C originate from the bright doubly-excited vibrational states, $\ket{E_b,1_j,1_k}$ and $\ket{E_b,2_j}$, created by multi-phonon transitions.

To identify the origin of the 2D peaks observed at $t_2=500\,{\rm fs}$, the reduced vibronic model in Section \ref{section_Markov_JL} is considered in 2D simulations where the three intrapigment vibrational modes are included in the system Hamiltonian and the remaining broad phonon spectrum $J_{\rm AR}(\omega)$ is approximately described by a Lindblad equation (Markovian noise). As shown in Fig.~\ref{FigS4_JL}(c), the 2D spectra at $t_2=0$ computed by the reduced model are well matched to the numerically exact 2D data computed by TEDOPA+MC, although a low-energy shoulder in the numerically exact results is absent in the reduced model results, as highlighted by white arrows in Fig.~\ref{FigS4_JL}(a) and (c). However, the reduced model cannot reproduce numerically exact 2D optical line shape at $t_2=500\,{\rm fs}$, as shown in Fig.~\ref{FigS4_JL}(d), where the cross peaks in the region B are absent in the reduced model results. It is found that the 2D peak intensities in the region A are enhanced when the vibrational damping rate of the three intrapigment modes is decreased in reduced model simulations (not shown here). This is in line with our interpretation of absorption peaks (see Fig.~\ref{FigS2_JL}(c)) where the three peaks in a vibrational sideband mainly originate from vibrationally hot dark excitons $\ket{E_d,1_k}$, which are weakly mixed with the vibrationally cold bright exciton state $\ket{E_b,0}$, making these vibronic eigenstates sensitive to the vibrational damping rate of the intrapigment modes. The differences in excitation and detection energies of the cross peaks in the region C are approximately two times larger than the vibrational frequencies of the intrapigment modes, as highlighted by a white arrow in Fig.~\ref{FigS4_JL}(d), indicating that these cross peaks originate from the bright doubly-excited vibrational states $\ket{E_b,1_j,1_k}$ and $\ket{E_b,2_j}$. Importantly, the absence of 2D cross peaks in the region B of Fig.~\ref{FigS4_JL}(d) suggests the possibility that these cross peaks originate from the vibrational excitations of the broad phonon spectrum $J_{\rm AR}(\omega)$.

To demonstrate that the vibrational excitations of the broad phonon spectrum are responsible for the cross peaks in the region B, we investigate 2D spectra computed by another reduced model in Section \ref{section_nonMarkov_JL} where the broad phonon spectral density $J_{\rm AR}(\omega)$ is modelled by a sum of 97 Lorentzian functions. For simplicity of analysis, we ignore the vibrational damping of the 100 effective modes and consider a phenomenological dephasing noise for the optical coherence dynamics during $t_1$ and $t_3$ by multiplying an exponential dephasing $\exp(-\gamma_{\rm ph}(t_1+t_3))$ to the third order molecular response function computed by using the reduced vibronic Hamiltonian $H_{\rm non-Markov}$ only. The phenomenological damping rate $\gamma_{\rm ph}$ is taken to be the uniform width of the 100 Lorentzian functions, $\sim (500\,{\rm fs})^{-1}$, considered in Fig.~\ref{FigS3_JL}(b). The reduced model results in the presence of vibrational damping of the 100 effective modes will be discussed later where the phenomenological dephasing is not considered ($\gamma_{\rm ph}=0$). In Fig.~\ref{FigS4_JL}(e), the 2D spectra at $t_2=0$ computed by using the reduced model Hamiltonian $H_{\rm non-Markov}$ are shown. The low-energy shoulder observed in TEDOPA+MC simulations is visible in the reduced model results, as highlighted by a white arrow. It is notable that for $t_2=500\,{\rm fs}$, the cross-peak structures in the regions B and C, observed in TEDOPA+MC simulations, are well reproduced by the reduced model, although the peak intensities in the region A are over-estimated by the reduced model, as shown in Fig.~\ref{FigS4_JL}(f), which is due to the absence of the vibrational damping of the intrapigment modes in simulations. It is found that the reduced model results obtained for double ($N=2$) and triple ($N=3$) vibrational excitation subspaces are qualitatively similar with minor quantitative differences in relative peak intensities (not shown here).

\begin{figure*}[t]
	\includegraphics[width=1\textwidth]{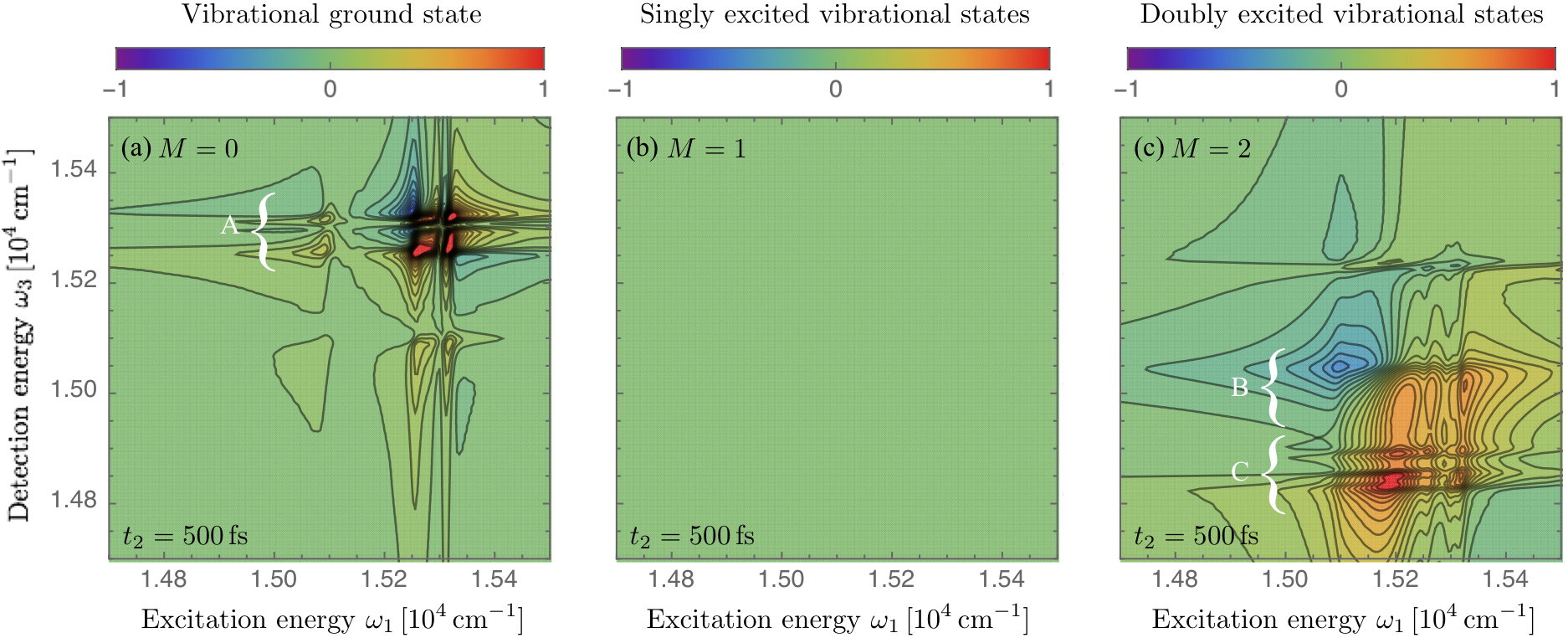}
	\caption{The SE component of rephasing 2D spectra conditional to the final vibrational states in the electronic ground state manifold. The contributions of (a) vibrational ground state $R_{\rm SE}^{(0)}$, (b) singly-excited vibrational states $R_{\rm SE}^{(1)}$, and (c) doubly-excited vibrational states $R_{\rm SE}^{(2)}$ to the total SE signal $R_{\rm SE}=\sum_{M=0}^{2}R_{\rm SE}^{(M)}$ are shown separately.}
	\label{FigS5_JL}
\end{figure*}

To clarify that the cross peaks in the regions B and C originate from the bright doubly-excited vibrational states $\ket{E_b,1_j,1_k}$ and $\ket{E_b,2_j}$, we consider third-order molecular response functions conditional to the final vibrational states in the electronic ground state manifold
\begin{equation}
	R_{\rm SE}^{(M)}(\omega_1,t_2,\omega_3) \propto \int_{0}^{\infty}dt_1 e^{-i\omega_1 t_1}\int_{0}^{\infty} dt_3 e^{i\omega_3 t_3} {\rm Tr}[I_{M}\mu^- \mathcal{U}(t_3)[\mathcal{U}(t_2)[\mu^+ \mathcal{U}(t_1)[\ket{g}\bra{g}\otimes\rho_{v}(T)\mu^-]]\mu^+]],
\end{equation}
where $I_M$ is an identity operator of the $M$ vibrational excitation sector, defined by $I_0=\ket{g,0}\bra{g,0}$, $I_1=\sum_{j=1}^{100}\ket{g,1_j}\bra{g,1_j}$ and $I_2=\sum_{j=1}^{100}\ket{g,2_j}\bra{g,2_j}+\sum_{j\neq k}\ket{g,1_j,1_k}\bra{g,1_j,1_k}$. The sum of these conditional response functions $R_{\rm SE}^{(M)}(\omega_1,t_2,\omega_3)$ is identical to the SE component of the rephasing spectra in Eq.~(\ref{eq:SE_JL}), shown in Fig.~\ref{FigS4_JL}(f). By analyzing $R_{\rm SE}^{(M)}(\omega_1,t_2,\omega_3)$ one can find which vibrational sector is responsible for the 2D peaks in the regions A, B, C. As shown in Fig.~\ref{FigS5_JL}(a), the 2D peaks in the region A originate from the vibrational ground state sector ($M=0$). The contribution of the singly-excited vibrational sector ($M=1$) is absent, as it is associated with optically dark states $\ket{E_d,1_j}$, as shown in Fig.~\ref{FigS5_JL}(b). The optically bright singly-excited vibrational states $\ket{E_b,1_j}$, which can contribute to 2D spectra, are populated by a vibrational damping of the bright doubly-excited vibrational states $\ket{E_b,1_j,1_k}$ and $\ket{E_b,2_j}$ in reduced model simulations, which is not considered in Fig.~\ref{FigS4_JL}(e,f) and Fig.~\ref{FigS5_JL}. The cross peaks in the regions B and C are induced by the doubly-excited vibrational sector ($M=2$), as shown in Fig.~\ref{FigS5_JL}(c), demonstrating that these 2D peaks originate from multi-phonon transitions.

\begin{figure*}[t]
	\includegraphics[width=1\textwidth]{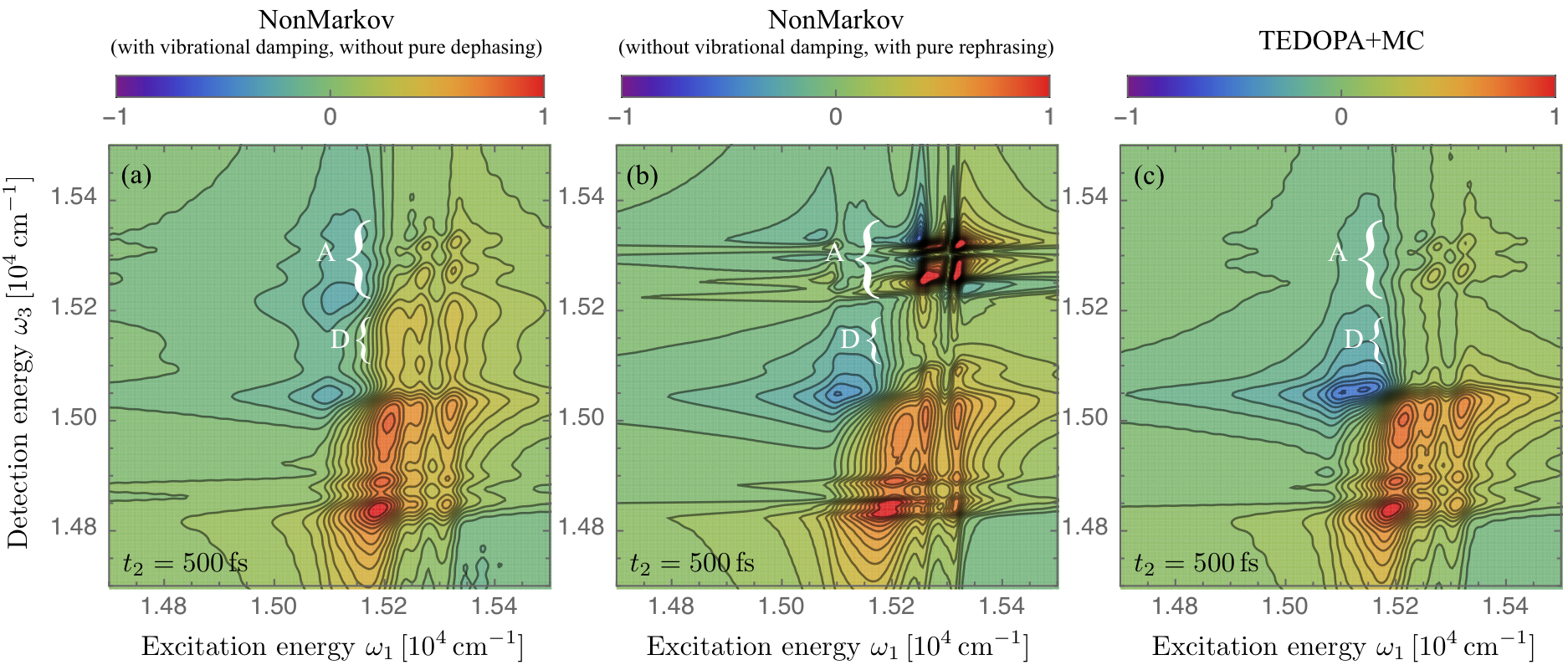}
	\caption{(a) The SE component of rephasing 2D spectra at $t_2=500\,{\rm fs}$ obtained by a reduced model where 100 damped effective modes are considered (non-Markovian noise, $\gamma_{\rm pd}=0$). (b) The reduced model results shown in Fig.~\ref{FigS4_JL}(f) where the vibrational damping is not considered, but phenomenological pure dephasing for $t_1$ and $t_3$ dynamics is considered ($\gamma_{\rm pd}\neq 0$). (c) Numerically exact 2D line shape obtained by TEDOPA+MC.}
	\label{FigS6_JL}
\end{figure*}

To clarify that the over-estimated peak intensities in the region A are due to the absence of the vibrational damping of the intrapigment modes, we consider the full reduced model in Section \ref{section_nonMarkov_JL} where the damping of the 100 effective modes is considered: the phenomenological dephasing rate for $t_1$ and $t_3$ dynamics is not considered here ($\gamma_{\rm pd}=0$). As shown in Fig.~\ref{FigS6_JL}(a), the peak intensities in the region A become weaker than those in the regions B and C when the vibrational damping is taken into account in reduced model simulations, similar to numerically exact 2D data obtained by TEDOPA+MC. The reduced model, however, introduces additional cross-peaks with weak intensities in a region D marked in Fig.~\ref{FigS6_JL}(a), which are absent in the reduced model results when the vibrational damping is not considered (see Fig.~\ref{FigS6_JL}(b)) as well as in the TEDOPA+MC results (see Fig.~\ref{FigS6_JL}(c)). These minor deviations between reduced and numerically exact results could be suppressed by considering a larger number of narrower Lorentzian functions in the fitting of the total spectral density $J(\omega)$, corresponding to a lower vibrational damping rate of the effective modes, or by fitting a bath correlation function instead of the total spectral density $J(\omega)$ to obtain more reliable parameters for reduced model simulations.

\section{Residual Environment} \label{sec:resenv}
\begin{figure*}[t]
	\includegraphics[width=1\textwidth]{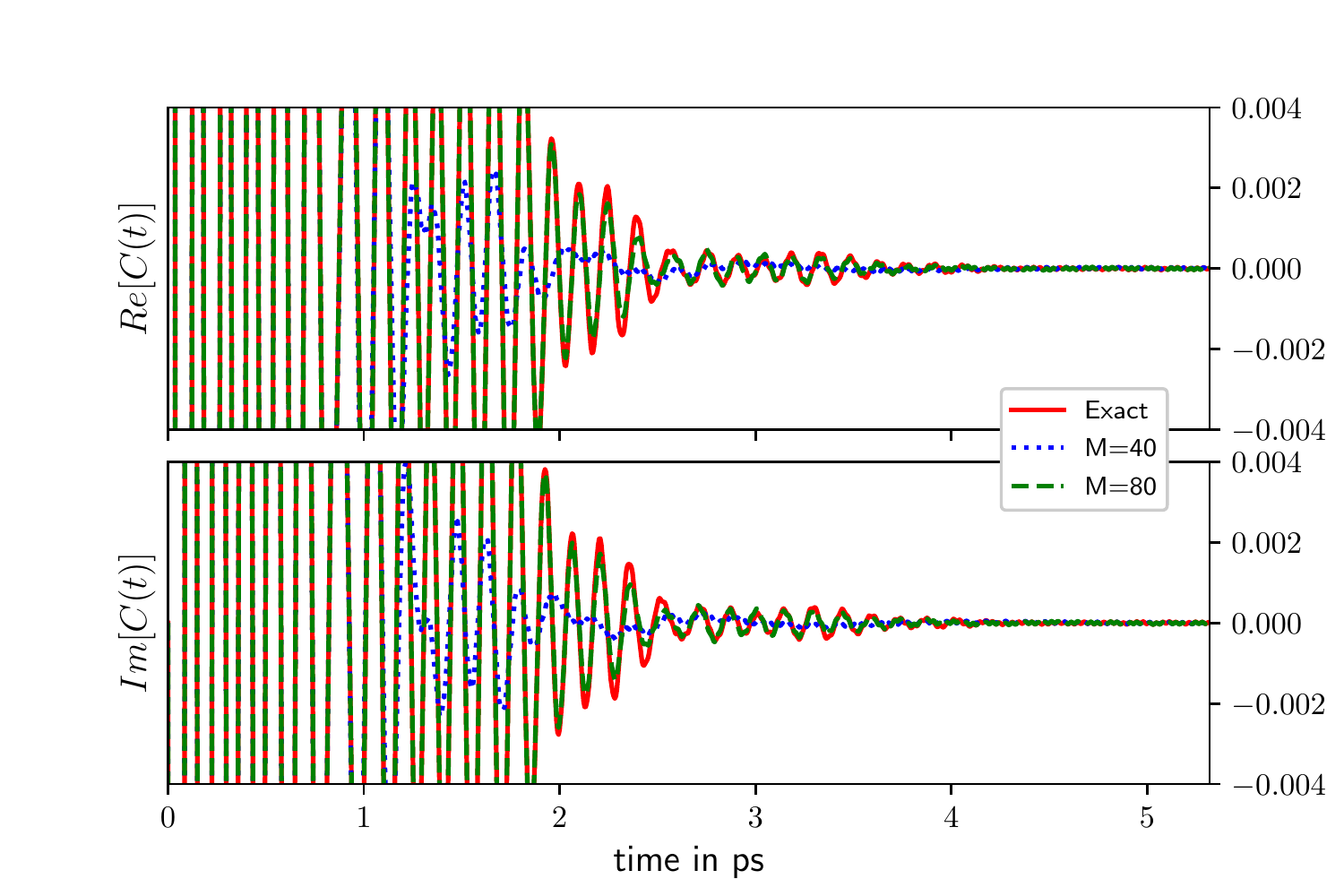}
	\caption{The real (top panel) and imaginary (bottom panel) part of the TTCF corresponding to a $D=400$ truncation of the WSCP TEDOPA chain. The exact TTCF (red solid line) is compared to the $M=40$ (dotted blue) and $M=80$ approximations. The ordinate axis has been rescaled as to have $Re[C(0)] = 1$.}
	\label{fig:TTCF_SI}
\end{figure*}

Here we provide an additional analysis of the error that is introduced by replacing the exact WSCP TEDOPA chain coefficients $\omega_n/k_n$ with the corresponding asymptotic values $\Omega/K$ provided in the main text (Eq.~(6)). To this end we first define the discrete (effective) spectra associated to the truncated TEDOPA chain. Once fixed an integer value $D>0$, we truncate the TEDOPA chain after $D$ sites. The truncated chain Hamiltonian therefore reads
\begin{align}
    H &= \sum_{n=1}^D \omega_n b_n^\dagger b_n + \sum_{n=1}^{D-1} \kappa_n (b_{n+1}^\dagger b_n + b_n^\dagger b_{n+1})  = \vec{b}^\dagger \widehat{H} \vec{b}, \\
     \vec{b} &= (b_1,b_2,\ldots,b_D)^T, \\
    \widehat{H} &= \left ( 
    \begin{array}{ccccc}
         \omega_1 &  \kappa_1 & 0 & \dots & 0 \\
         \kappa_1 &  \omega_2 & \kappa_2 & \dots &0 \\
         \dots&  \dots & \dots & \dots & \dots \\
         0 &  \dots& \dots & \kappa_{D-1} &\omega_D
    \end{array}
    \right).
\end{align}
If we denote by $S = \text{diag}(e_1,e_2,\ldots,e_D)$ with $j=1,2,\ldots,D$ the eigenvalues of the normal modes of $\widehat{H}$ and by $U$ the unitary transformation such that $\widehat{H} = U^\dagger S U$, the discrete spectral density truncated chain can be written as
\begin{equation}
J^D(\omega) = \kappa_0^2 \sum_{j=1}^D \left |U_{1,j}\right | ^2  \delta(\omega - e_j),) = \kappa_0^2 \sum_{j=1}^D  w_j  \delta(\omega - e_j),
\end{equation}
where $k_0$ is the coupling coefficient between the system and the first chain site, $\delta(x)$ is the Dirac-$\delta$ function, and $ w_j =  \left | U_{1,j} \right |^2$. The spectral density $J_D(\omega)$ determines the two-time correlation function  (TTCF) $C_D(t)$ as
\begin{equation}
C_D(t) = \frac{\kappa_0^2}{\pi} \sum_{j=1}^D w_j \exp\left ({-i t e_j} \right ).
\end{equation}
For any fixed value of $D$, the quality of the approximation of the exact TEDOPA chain can be therefore assessed also by looking at the deviations between the TTCF corresponding to the exact chain coefficients and the approximating chain where we have operated the replacements $\omega_n \to \Omega$ and $\kappa_n \to K$ for $M<n\leq D$. Figure~\ref{fig:TTCF_SI} shows, for the same truncation value $D=400$ used in Fig. 2(b) of the main text, the evolution of the TTCF determined by the exact TEDOPA chain, and the TTCF corresponding to $M=40$ and $M=80$ approximations. In order to better understand the relative deviation of the approximated TTCFs from the exact one, we set $\kappa_0/\pi = 1$, so that $Re[C_D(0)] = 1$ ($Im[C_D(0)] = 0$ by definition). We see that for $M=80$ there is an excellent agreement between the TTCFs determined by the exact and by the approximated chain, whereas the error becomes much larger for $M=40$. The comparison between the different TTCFs allows moreover to appreciate the earlier onset of deviations for $M=40$ w.r.t. the $M=80$ case.  
\section{Markovian closure: derivation of the parameters} \label{sec:TSO}
In this section we provide full detail on the derivation of the universal Markovian Closure (MC) parameters. The numerical values of the parameters for sizes $N=6,8$ and $10$ of the MC are reported in Table \ref{tab:summary_closure_parameters_N6}, \ref{tab:summary_closure_parameters_N8}, and \ref{tab:summary_closure_parameters_N10}.

\subsection{Transformation to surrogate oscillators and fitting procedure}
In the first step of the derivation we rewrite the interaction Hamiltonian between the primary environment and the residual environment
\begin{align} \label{eq:h_smr}
    H_\text{PE,R}  = K (b_{M+1}^\dagger b_{M} + b_{M+1} b_{M}^\dagger)
\end{align}
(see Eq.~(8) of the main text) in terms of the normal modes $\tilde{b}_\omega$ of the residual environment Hamiltonian
\begin{align}
    H_R &= \Omega \hspace{-7pt}\sum_{m=M+1}^{\infty} b_m^\dagger b_m   
+ K \hspace{-7pt}\sum_{m=M+1}^{\infty}
(b_{m+1}^\dagger b_m + b_m^\dagger b_{m+1}), \label{eq:truncApproxHam}
\end{align}
as
\begin{align}
    H_\text{SM,R} &= \int_0^{2\Omega} \d \omega\; h(\omega)  (b_M \tilde{b}_\omega^\dagger + b_M^\dagger \tilde{b}_\omega) 
    = \int_0^{2\Omega} \d \omega\; h(\omega)  (x_M \tilde{x}_\omega + p_M \tilde{p}_\omega), \label{eq:hhh}
\end{align}
with $h^2(\omega) = J_\infty(\omega)/ \pi$ and 
\begin{equation}
    J_\infty(\omega) = \frac{\sqrt{(\omega-\omega_\text{min})(\omega_\text{max}-\omega)}}{2}, \label{eq:asymptoticResidual}
\end{equation}
(see Eq.~(10) of the main text).
Our aim is now to find an auxiliary system made up of a (small) {discrete set} of bosonic modes which, themselves, are undergoing a Lindbladian dynamics that, upon interaction with the system and primary environment, induces on the latter the same reduced {evolution (for one- and multi-time quantities)}
as the residual environment. 
More specifically, we need to determine an auxiliary system  whose free semigroup dynamics generated by $\mathcal{L}_\text{aux}(\cdot)$ starting from a Gaussian initial state $\rho_\text{aux}(0)$ and interacting with the extended system via 
\begin{align}
    H_\text{SM,aux} = \sum_{k=1,2} A_{\text{SM},k}\otimes B_{\text{aux,k}}
\end{align}
satisfies (approximately) the identity
\begin{align} \label{eq:equivalence_constraints}
    \braket{A_{\text{R}, k}(t) A_{\text{R}, j}(0)} = \braket{B_{\text{aux}, k}(t) B_{\text{aux}, j}(0)}
\end{align}
for all $k,j \in \lbrace 1,2 \rbrace$ and for all $t > 0$.
Here, $A_{\text{SM},k}$ denote the Hermitian coupling operators of the system (see \cref{eq:hhh}),
{\begin{align}
	A_{\text{SM}, 1} &:= x_M \qquad	A_{\text{SM}, 2}:= p_M,
\end{align}
$A_{\text{R},k}$ denote the coupling operators of the residual environment,
\begin{align}
    A_{\text{R}, 1} &:= \int_{0}^{2\Omega} \d \omega \; h(\omega) \tilde{x}_\omega,
    \qquad A_{\text{R}, 2} := \int_{0}^{2\Omega} \d \omega \; h(\omega) \tilde{p}_\omega,
\end{align}
and 
\begin{align}
        B_{\text{aux}, 1} &:= \sum_{n = 1}^{N} \frac{1}{\sqrt{2}}(c_{n} d_n + c_{n}^* d_n^\dagger), 
       \qquad B_{\text{aux}, 2} :=\sum_{n = 1}^{N} \frac{\ii}{\sqrt{2}}(c_{n}^* d_n^\dagger - c_{n} d_n)
\end{align}
denote the coupling operators of the auxiliary environment.}


In order to construct explicitly a well-defined auxiliary system, we follow the Transformation to Surrogate Oscillators (TSO) approach described in \cite{mascherpa20}. Our aim is thus to find a proper set of oscillators governed by the nearest-neighbor Hamiltonian
\begin{align}
    H_{\text{aux}} &:= \sum_{n = 1}^N \Omega_n d_n^\dagger d_n + \sum_{n = 1}^{N-1} g_n (d_{n} d_{n+1}^\dagger + d_{n}^\dagger d_{n+1})
    \label{eq:tso_hamiltonian}
\end{align}
and subject to the local Lindblad dissipator
\begin{align}
    \mathcal{D}(\rho) := \sum_{n = 1}^{N} \Gamma_n \left(d_n \rho d_n^\dagger - \frac{1}{2} \{d_n^\dagger d_n, \rho\} \right).
    \label{eq:tso_dissipator}
\end{align}
{For} the given asymptotic residual spectral density \cref{eq:asymptoticResidual}, the correlation matrix reads
\begin{align}
 	\mathcal{C}_{\infty}(t) &= 
  	\begin{pmatrix}
 		\braket{A_{\text{R}, 1}(t) A_{\text{R}, 1}(0)} & \braket{A_{\text{R}, 1}(t) A_{\text{R}, 2}(0)} \\
  		\braket{A_{\text{R}, 2}(t) A_{\text{R}, 1}(0)} & \braket{A_{\text{R}, 2}(t) A_{\text{R}, 2}(0)}
 	\end{pmatrix} \notag
	\\
 	&= \frac{1}{2\pi} \int_{\omega_{\min}}^{\omega_{\max}} \d \omega \; J_\infty(\omega) \e^{-\ii \omega t}
 	\begin{pmatrix}
 		1  & \ii \\
 		-\ii & 1
 	\end{pmatrix} 
 	= \frac{1}{2} C_{\infty}(t)
 	\begin{pmatrix}
 		1  & \ii \\
 		-\ii & 1
 	\end{pmatrix},
\end{align}
where
\begin{align}
    C_{\infty}(t) &:= \frac{1}{\pi} \int_{\omega_{\min}}^{\omega_{\max}} \d \omega \; J_\infty(\omega) \e^{-\ii \omega t}. \label{eq:general_correlation_function}
\end{align}
Using the linear transformation
\begin{align}
    \omega: [-1,1] \rightarrow [\omega_{\min}, \omega_{\max}], \; \omega(x) := 2\kappa x + \Omega,
\end{align}
with $\Omega$ and $K$ defined by
\begin{align} 
    \omega_n &\xrightarrow[n \to \infty] { } \frac{\omega_\text{min}+\omega_\text{max}}{2} \stackrel{\text{def}}{=} \Omega,  \nonumber \\
    \kappa_n &\xrightarrow[n \to \infty] { } \frac{\omega_\text{max}-\omega_\text{min}}{4} \stackrel{\text{def}}{=} K \label{eq:limCoeff}
\end{align}
(see Eq.~(6) of the main text), we are able to transform \cref{eq:general_correlation_function} into
\begin{align}
    C_{\infty}(t) &= \frac{1}{\pi} \int_{\omega_{\min}}^{\omega_{\max}} \d \omega \; J_{\infty}(\omega) \; \e^{-\ii \omega t} \notag\\
    &= \frac{1}{\pi} \int_{-1}^{1} \d x \; 2K^2 \, j_{\text{sc}}(x) \; \e^{-\ii(2K x + \Omega) t} 
    = K^2 \; \e^{-\ii \Omega t} C_{\text{sc}}(2K t).
    \label{eq:relation_asymptotic_semi_circle_cft}
\end{align}
Here, we introduced the symmetric semi-circle spectral density
\begin{align}
    j_{\text{sc}} : [-1,1] \rightarrow [0,\infty), \; j_\text{sc}(\omega) := \sqrt{1 - \omega^2} \label{eq:sd_symmetric_semi_circle}
\end{align}
as well as the correlation function
\begin{align}
    C_{\text{sc}}(t) &:= \frac{2}{\pi} \int_{-1}^{1} \d \omega \; j_{\text{sc}}(\omega)  \;  \e^{-\ii \omega t}
    = \frac{2J_1(t)}{t} = J_0(t) + J_2(t),
    \label{eq:semi_circle_correlation_function}
\end{align}
where $J_0, J_1$ and $J_2$ are Bessel functions of the first kind.
Hence, any asymptotic correlation function $C_{\infty}$ can be recovered from the generic correlation function $C_{\text{sc}}$ by rescaling with $\frac{K^2}{2}$ and multiplication with the time-dependent phase factor $\e^{-\ii \Omega t}$.
Summarizing, having fixed the structure of the auxiliary environment as in \cref{eq:tso_hamiltonian,eq:tso_dissipator}, our aim is now to determine the TSO coefficients $N$ and the coefficients $c_n, \Omega_n, g_n$ so that the relations in \cref{eq:equivalence_constraints} hold {approximately, to a degree which we can fix a-priori}.

To do so, we proceed in two steps.
In the first step, we determine an exponential sum approximation of $C_{\text{sc}}$,
\begin{align}
    C_{\text{sc}}(t) &\approx \sum_{k = 1}^{N} w_k \e^{\lambda_k t}, \quad w_k, \lambda_k \in \C, \label{eq:prony_exponential_sum}
\end{align}
by means of the approximated Prony method~\cite{beylkin2005}.
Here, we can either specify the number of terms $N$ directly or the desired accuracy in norm which then dictates $N$.
Since, in general, the weights $w_k$ are indeed complex-valued or partially negative, the corresponding surrogate environment
{made up of non-interacting pseudomodes} is shown to be ill-defined~\cite{garraway97}.
Hence, in the second step, we follow Ref.~\cite{mascherpa20} to construct a well-defined set of coupled oscillators.

The auxiliary environment correlation matrix is given by
\begin{align}
    \mathcal{C}_{\text{aux}}(t) &= 
	\begin{pmatrix}
		\braket{B_{\text{aux}, 1}(t) B_{\text{aux}, 1}(0)} & \braket{B_{\text{aux}, 1}(t) B_{\text{aux}, 2}(0)} \\
		\braket{B_{\text{aux}, 2}(t) B_{\text{aux}, 1}(0)} & \braket{B_{\text{aux}, 2}(t) B_{\text{aux}, 2}(0)}
	\end{pmatrix} \notag\\
	&= 
	\frac{1}{2} \sum_{m,n = 1}^{N} c_m c_n^* \braket{d_m(t) d_n^\dagger(0)}
	\begin{pmatrix}
	    1    & \ii  \\
	    -\ii & 1
	\end{pmatrix} 
	= \frac{1}{2} C_{\text{aux}}(t) 
	\begin{pmatrix}
	    1    & \ii  \\
	    -\ii & 1
	\end{pmatrix},
\end{align}
where we defined the correlation function 
\begin{align}
    C_{\text{aux}}(t) &:= \sum_{m,n = 1}^{N} c_m c_n^* \braket{d_m(t) d_n^\dagger(0)}.
\end{align}
{As show in detail in Ref.~\cite{mascherpa20}, $C_{\text{aux}}$ can be rephrased as}
\begin{align}
    C_{\text{aux}}(t) &= \sum_{m,n = 1}^N c_m c_n^* (\e^{M t})_{m,n}
\end{align}
where
\begin{align}
    M &:= 
    \begin{pmatrix}
        \alpha_1 & \beta_1  &          & \\
        \beta_1  & \alpha_2 & \beta_2  & \\
                 & \beta_2  & \alpha_3 & \\
                 &          &          & \ddots
    \end{pmatrix}
\end{align}
and $\alpha_n := -\frac{\Gamma_n}{2} - \ii \Omega_n$ and $\beta_n := -\ii g_n$.
The remaining, very challenging task is to fit the parameters $\{\Omega_n\}_{n = 1}^N$, $\{g_n\}_{n = 1}^{N-1}$, $\{\Gamma_n\}_{n = 1}^N$ and $\{c_n\}_{n = 1}^N$ such that
\begin{align}
    \|C_{\text{sc}}(t) - C_{\text{aux}}(t) \| < \epsilon \quad \forall t \in [0, T_f]
\end{align}
for some desired $\epsilon, T_f > 0$.
Since $M$ is a complex-valued symmetric matrix, i.e., $M^T = M$, there exists $U \in \C$ such that $U^T U = \openone$ and
\begin{align}
     U^T M U &= \diag{\lambda_1^\prime, \dots, \lambda_N^\prime} =: \Lambda.
\end{align}
Using this equivalence, the optimization problem decomposes into two coupled subproblems: Find $\{\Omega_n\}_{n = 1}^N$, $\{g_n\}_{n = 1}^{N-1}$, $\{\Gamma_n\}_{n = 1}^N$ and $\{c_n\}_{n = 1}^N$ such that
\begin{align}
    &\sum_{n = 1}^{N} |\lambda_n - \lambda_n^\prime| \rightarrow \min \label{eq:eigenvalue_optimization}\\
    &\sum_{n = 1}^{N} |w_n - w_n^\prime| \rightarrow \min \label{eq:weight_optimization}
\end{align}
where $w_n^\prime := \sum_{k, \ell = 1}^{N} c_k c_\ell^* U_{nk} U_{n\ell}$.
In contrast to Ref.~\cite{mascherpa20} we do not solve these two subproblems separately but rather combine both equations into a single cost function.
This approach is motivated by the fact that while a solution to \cref{eq:eigenvalue_optimization} is very easily found, imposing this solution on the optimisation problem of \cref{eq:weight_optimization} might lead to local minima.
Hence, it makes sense to combine both optimization problems into a single cost function,
\begin{align}
    \mathcal{J} &:=\sum_{n = 1}^N |\lambda_n - \lambda^\prime_n| + |w_n - w_n^\prime| \rightarrow \min.
    \label{eq:combined_cost_function}
\end{align}
Once we have found a surrogate environment mimicking $C_{\text{sc}}$, the approximation of any asymptotic correlation function $C_{\infty}$ is easily recovered from \cref{eq:relation_asymptotic_semi_circle_cft}.
In particular, we find
\begin{align}
    C_{\infty}(t) &= K^2 \e^{-\ii \Omega t} C_{\text{sc}}(2K t) 
    \approx K^2 \e^{-\ii \Omega t} C_{\text{aux}}(2K t) 
    = K^2 \e^{-\ii \Omega t} \sum_{m, n = 1}^{N} c_m c_n^* (\e^{M 2 K t})_{mn} 
    = \sum_{m, n = 1}^{N} (K c_m ) (K c_n^*) \left(\e^{(2K M - \ii \Omega )t}\right)_{mn}.
\end{align}
Hence, we can identify the parameters of the surrogate environment corresponding to $C_{\infty}$ as
\begin{align}
    \begin{split}
        \Omega_n^\prime &:= 2 K \Omega_n + \Omega, \\
        \Gamma_n^\prime &:= 2 K \Gamma_n, \\
        g_n^\prime &:= 2 K g_n, \\
        c_n^\prime &:= K c_n      
    \end{split}
\end{align}
for all $n$.

\subsection{Fitting results}

\begin{figure*}[h]
    \centering
    \includegraphics[width=\linewidth]{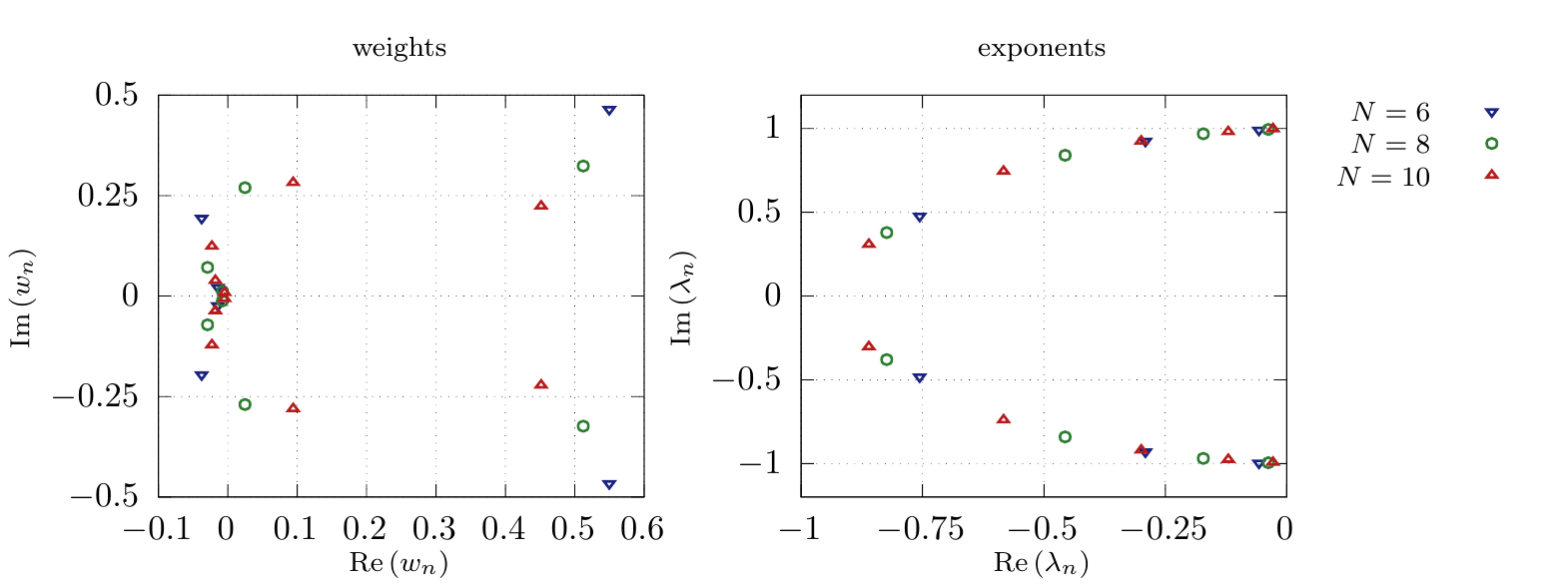}
    \caption{\textbf{Prony parameters:} We depict the complex weights $w_n$ and exponents $\lambda_n$ of \cref{eq:prony_exponential_sum} obtained through a Prony analysis of the correlation function defined in \cref{eq:semi_circle_correlation_function}.}
    \label{fig:prony_parameters}
\end{figure*}

In the following we briefly summarize the solutions to \cref{eq:eigenvalue_optimization,eq:weight_optimization} for the correlation function in \cref{eq:semi_circle_correlation_function} and a total number of $N = 6, 8$ and $10$ surrogate modes.
Completing the exponential sum approximation in \cref{eq:prony_exponential_sum} we find the weights $\{w_n\}_{n=1}^{N}$ and exponents $\{\lambda_n \}_{n=1}^{N}$ depicted in \cref{fig:prony_parameters}.

\begin{figure*}[h]
    \centering
    \includegraphics[width=\linewidth]{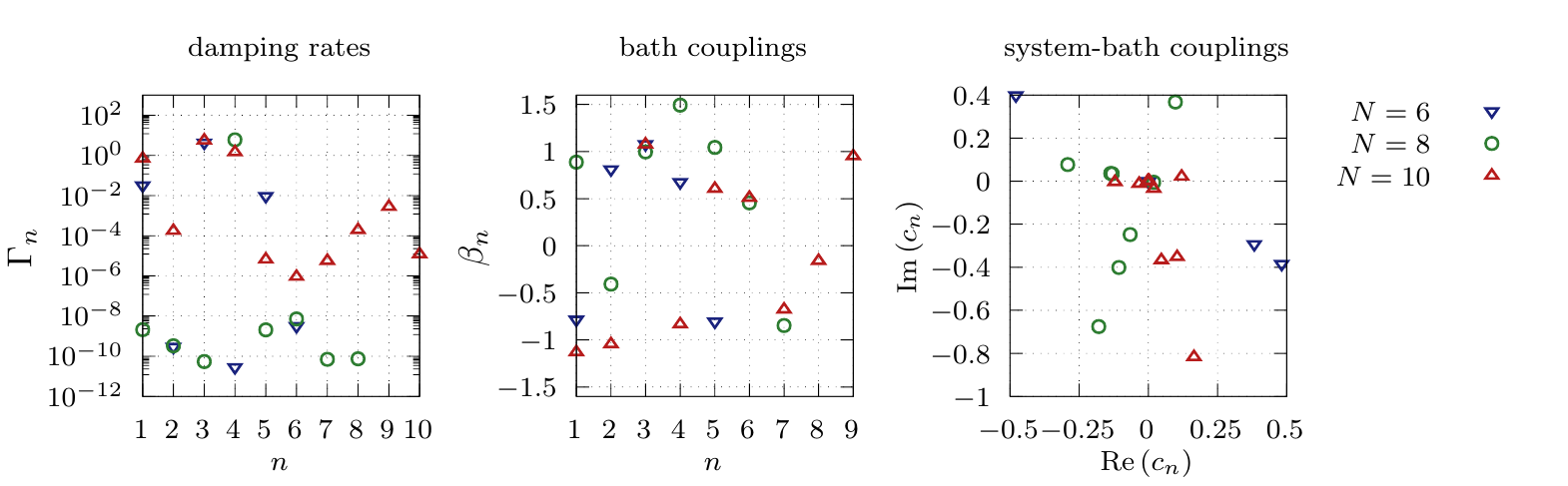}
    \caption{\textbf{TSO parameters:~}
    We show the fitted damping rates, intra bath couplings, and system-bath couplings.
    Note that for all three closures only very few oscillators are significantly damped.
    }
    \label{fig:tso_parameters}
\end{figure*}

\begin{figure*}[h]
    \centering
    \includegraphics[width=1.0\linewidth]{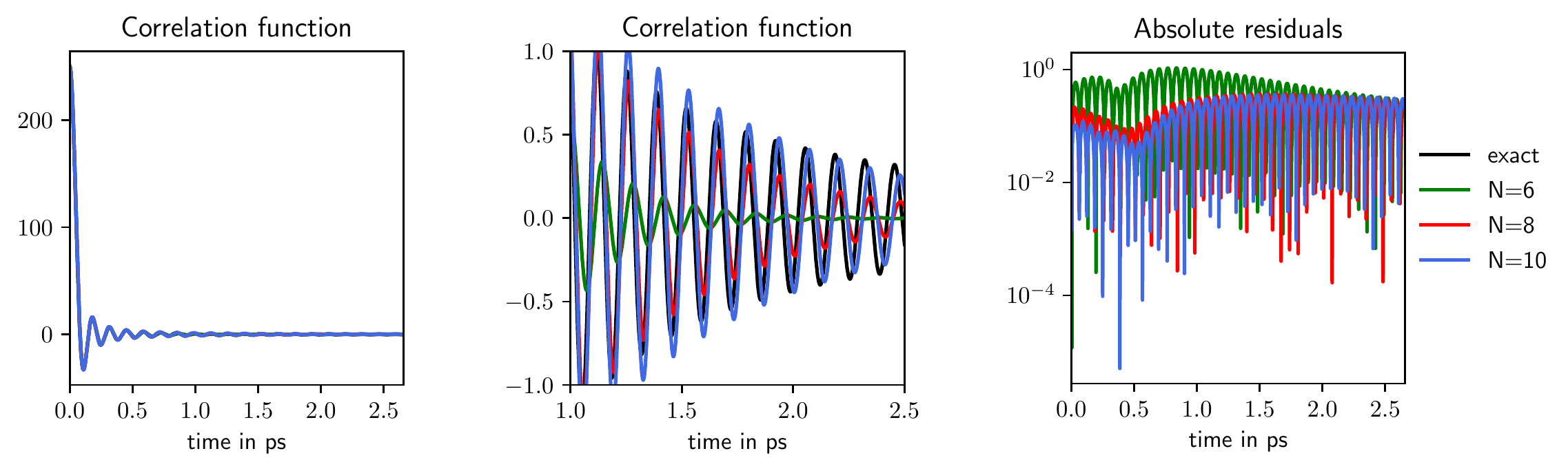}
    \caption{\textbf{Approximation of correlation function:} TSO fit of correlation function $C_{\mathrm{sc}}$ defined in \cref{eq:semi_circle_correlation_function}. The left and center panels compare the exact correlation function and the approximations provided by $N=6,8$ and $10$ surrogate modes. The right panel shows the absolute residuals for the different approximations.}
    \label{fig:fit_correlation_function}
\end{figure*}

\begin{figure*}[htpb]
    \centering
    \includegraphics[width=\linewidth]{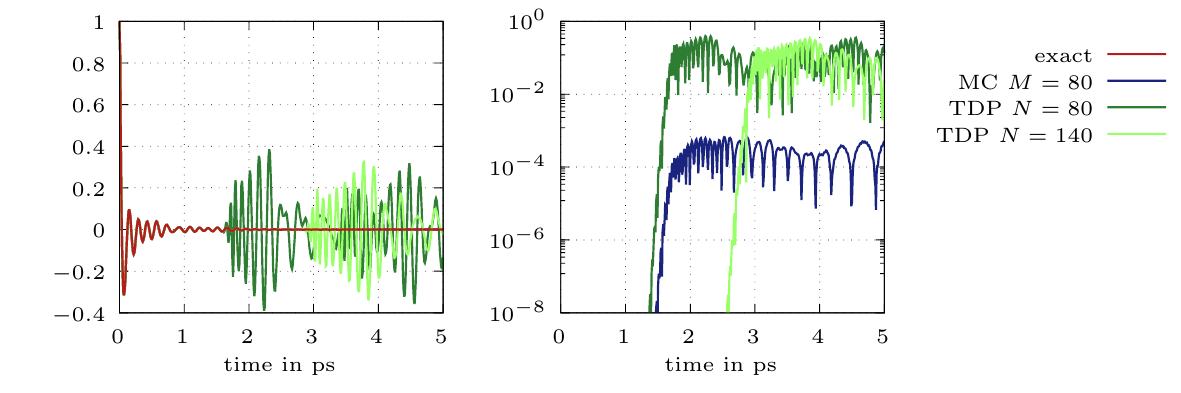}
    \caption{\textbf{Comparison of Two-time correlation functions for WSCP example:} The WSCP exact two-time correlation function is compared to those obtained by truncating the TEDOPA chain after $N=80,140$ sites and the one corresponding to a TEDOPA+MC of $N=80+8$ sites, namely with an MC of $8$ surrogate modes applied at the $M=80$ chain site. 
    }
    \label{fig:comparison_ttcf_wscp}
\end{figure*}

Interestingly, both the weights and the exponents come in conjugate pairs.
In principle, one could think of grouping these conjugate pairs and construct a surrogate environment for each of them.
However, this is not feasible in general since the weights do not have positive real part.
Instead one has to divide the weights and exponents into larger groups such that the sum of weights has positive real part for each group.
In this work, we decided to not further subdivide the parameters but rather take them as a single group.
As a consequence, the number of parameters to be optimized remains quite high.
More precisely, \cref{eq:eigenvalue_optimization,eq:weight_optimization} involve $4N-1$ optimization parameters, i.e., for $N = 6,8,10$ we have around $20$ to $40$ unknowns.
In fact, this large number of unknowns in combination with the non-linearity of the problem makes solving \cref{eq:eigenvalue_optimization,eq:weight_optimization} very challenging.
Hence, in order to simplify the optimization slightly we constrain all the surrogate frequencies to vanish, i.e., $\Omega_n = 0$ for all $n$.
This can be motivated in two ways.
Firstly, when mapping the symmetric semi-circle spectral density to a chain via TEDOPA we would indeed arrive at a chain that has vanishing surrogate frequencies for all modes.
Secondly, when considering only two conjugate modes of the exponential sum approximation the analytical formulas derived in \cite{garraway97,mascherpa20} would also lead to two surrogate modes with vanishing frequencies. 
Thus, it is reasonable to speculate that even for a larger number of conjugate pairs an optimal solution satisfies $\Omega_n = 0$ for all $n$.
\cref{tab:summary_closure_parameters_N6,tab:summary_closure_parameters_N8,tab:summary_closure_parameters_N10} and \cref{fig:tso_parameters} summarize the optimal parameters found for $N = 6,8$ and $10$.
In contrast to the weights and exponents obtained by the approximated Prony method, there seems to be no particular structure in the distribution of the parameters.
However, as \cref{fig:fit_correlation_function} clearly shows, we have indeed found surrogate environments of different size that mimic the initial correlation function in \cref{eq:semi_circle_correlation_function} extremely well.
Additionally, Fig. \ref{fig:comparison_ttcf_wscp} compares the effective spectral density to the exact one.
Here, by effective spectral density we mean the Fourier transform of the two-time correlation function of the auxiliary environment.
Naturally, a non-unitary Lindblad evolution only defines $C_{\text{aux}}$ for positive times $t$.
In order to compute the Fourier transform, we thus extend the correlation function $C_{\text{aux}}$ into the negative time domain by defining
\begin{align}
    \overline{C}_{\text{aux}}: \R \rightarrow \C, \; \overline{C}_{\text{aux}}(t) := 
    \begin{cases}
        C_{\text{aux}}(t)   & \text{for } t \geq 0\\
        C_{\text{aux}}^*(-t) & \text{for } t < 0
    \end{cases}.
\end{align}
Hence, we can identify a real-valued spectral density of the auxiliary environment, $j_{\text{aux}}: \R \rightarrow [0,\infty)$,
\begin{align}
    j_{\text{aux}}(\omega) &:= \frac{1}{4} \int_{-\infty}^{\infty}  \d t \; \overline{C}_{\text{aux}}(t) \, \e^{\ii \omega t} 
    = \frac{1}{4} \int_{-\infty}^{\infty} \d t\; \Re{\left(\overline{C}_{\text{aux}}(t) \, \e^{\ii \omega t}\right)}.
\end{align}

\begin{table*}[htbp]
    \centering
    \caption{\textbf{Summary of closure parameters for $N = 6$}}
    \label{tab:summary_closure_parameters_N6}
    \begin{tabular}{c | rc rc rc rc rc c}
        $n$ & 1 && 2 && 3 && 4 && 5 && 6 \\ \hline
         $\Gamma_n$ & $-1.60\cdot 10^{-2}$ && $-1.48\cdot 10^{-10}$ && $-2.18\cdot 10^0$ && $-1.44 \cdot 10^{-11}$ && $-4.79 \cdot 10^{-3}$ && $-1.57 \cdot 10^{-9}$  \\
         $\Re{c_n}$ &  $2.74\cdot 10^{-5}$  && $-4.79 \cdot 10^{-1}$  && $6.34\cdot 10^{-6}$  &&  $4.82 \cdot 10^{-1}$  && $-1.40 \cdot 10^{-6}$ && $3.83 \cdot 10^{-1}$\\
         $\Im{c_n}$ &  $-1.11\cdot 10^{-5}$  && $3.99 \cdot 10^{-1}$  && $-3.53\cdot 10^{-6}$  &&  $-3.84 \cdot 10^{-1}$ && $2.45 \cdot 10^{-6}$ && $-2.93 \cdot 10^{-1}$\\
         $g_n$ & & $0.79$ && $-0.813$ && $-1.08$ && $-0.68$ && $0.81$ & 
    \end{tabular}
\end{table*}

\begin{table*}[htbp]
    \centering
    \caption{\textbf{Summary of closure parameters for $N = 8$}}
    \label{tab:summary_closure_parameters_N8}
    \begin{tabular}{c | rc rc rc rc rc}
        $n$ & 1 && 2 && 3 && 4 && 5 & \\ \hline
         $\Gamma_n$ & $-1.06\cdot 10^{-9}$ && $-1.64\cdot 10^{-10}$ && $-2.70\cdot 10^{-11}$ && $-2.98\cdot 10^{+0}$ && $-1.02\cdot 10^{-9}$ & \\
         $\Re{c_n}$ &  $-6.58 \cdot 10^{-2}$  && $-1.31 \cdot 10^{-1}$ && $-1.79 \cdot 10^{-1}$  && $1.92 \cdot 10^{-2}$   && $9.77 \cdot 10^{-2}$ & \\
         $\Im{c_n}$ &  $-2.48 \cdot 10^{-1}$  &&  $3.47 \cdot 10^{-2}$ && $-6.75 \cdot 10^{-1}$  &&  $-5.08 \cdot 10^{-3}$ && $3.68 \cdot 10^{-1}$ &  \\
         $g_n$ & & $-0.89$ && $0.41$ && $-1.00$ && $-1.49$ && $-1.04$ \\[5pt]
         $n$ & 6 && 7 && 8 &&  &&   \\ \hline
         $\Gamma_n$ & $-3.61\cdot 10^{-9}$ && $-3.53\cdot 10^{-11}$ && $-3.73\cdot 10^{-11}$ &&  &&  \\
         $\Re{c_n}$ &  $-1.36 \cdot 10^{-1}$  && $-1.06 \cdot 10^{-1}$  &&  $-2.91 \cdot 10^{-1}$ &&  &&  \\
         $\Im{c_n}$ &  $3.60 \cdot 10^{-2}$  && $-4.01 \cdot 10^{-1}$  &&  $7.73 \cdot 10^{-2}$ &&  &&  \\
         $g_n$ & & $-0.45$ && $0.85$ && && 
    \end{tabular}
\end{table*}

\begin{table*}[htbp]
    \centering
    \caption{\textbf{Summary of closure parameters for $N = 10$}}
    \label{tab:summary_closure_parameters_N10}
    \begin{tabular}{c | rc rc rc rc rc}
        $n$ & 1 && 2 && 3 && 4 && 5 & \\ \hline
         $\Gamma_n$ & $-3.43 \cdot 10^{-1}$ && $-8.67 \cdot 10^{-5}$ && $-2.73 \cdot 10^{+0}$ && $-7.09 \cdot 10^{-1}$ && $-3.24 \cdot 10^{-6}$ & \\
         $\Re{c_n}$ & $-1.32 \cdot 10^{-3}$ && $3.32 \cdot 10^{-3}$  && $-2.40 \cdot 10^{-3}$  && $1.94 \cdot 10^{-2}$  && $-3.32 \cdot 10^{-2}$ & \\
         $\Im{c_n}$ &  $4.62 \cdot 10^{-4}$ && $5.49 \cdot 10^{-4}$ && $-1.48 \cdot 10^{-3}$ && $-3.55 \cdot 10^{-2}$ && $-1.20 \cdot 10^{-2}$  &  \\
         $g_n$ & & $1.13$ && $1.05$ && $-1.08$ && $0.83$  && $-0.60$ \\[5pt]
         $n$ & 6 && 7 && 8 && 9 && 10  \\ \hline
         $\Gamma_n$ & $-4.50 \cdot 10^{-7}$ && $-2.79 \cdot 10^{-6}$ && $-9.48 \cdot 10^{-5}$ && $-1.37 \cdot 10^{-3}$ && $-5.95 \cdot 10^{-6}$ \\
         $\Re{c_n}$ &  $1.04 \cdot 10^{-1}$ && $1.21 \cdot 10^{-1}$  && $1.65 \cdot 10^{-1}$  && $-1.21 \cdot 10^{-1}$ && $4.72 \cdot 10^{-2}$ \\
         $\Im{c_n}$ &  $-3.53 \cdot 10^{-1}$ && $2.08 \cdot 10^{-2}$ && $-8.17 \cdot 10^{-1}$ && $-4.45 \cdot 10^{-3}$ && $-3.67 \cdot 10^{-1}$   \\
         $g_n$ & & $-0.51$ && $0.68$ && $0.16$ &&  $-0.95$
    \end{tabular}
\end{table*}

%
%
%
%
\section{TEDOPA with Markovian Closure}\label{sec:twm}
Here we provide details on the TEDOPA+MC simulations used to derive the  2DES spectra discussed in the previous sections.

\subsection{Computation of the third-order response function}
\label{sec:computation_third_order_response}

For the equilibrium state $\rho_{\text{eq}} = \ket{g, 0}\bra{g, 0}$, where $\ket{g}$ is the ground state of the electronic Hamiltonian and $\ket{0}$ the vacuum state of the environment, the third-order response function for the stimulated emission pathway is the argument of the inner integral of \cref{eq:SE_JL}~\cite{May2011}, which we rewrite as
\begin{align}
   \hspace{-15pt}R_{\text{se}} &= \Tr \left\lbrace 
   \mu^- \mathcal{U}(t_3)\Big[\mathcal{U}(t_2)\big[\mu^+ \mathcal{U}(t_1)[\ket{g, 0}\bra{g, 0}\mu^-]\big]\mu^+\Big]
   \right\rbrace,
    \label{eq:stimulated_emission_response_function}
\end{align}
where recall that $\mathcal{U}(t)[\cdot] = e^{-i H t}\cdot e^{i H t}$ denotes the unitary propagator of the closed system and 
\begin{align}
    \begin{split}
        \hat{\mu}_+ &:= \ket{e_1} \bra{g} + \ket{e_2} \bra{g}  \\
        \hat{\mu}_- &:= \ket{g} \bra{e_1} + \ket{g} \bra{e_2}
    \end{split}
\end{align}
denote the components of the normalized dipole operator.

Now, extending the results of \cite{tamascelli18} to the case of a multi-time object, since we are dealing
with a Gaussian environment we can replace the unitary evolution by the dissipative one fixed by the Lindblad
equation given by the surrogate Harmonic oscillators defined in Sec.\ref{sec:TSO};
the general proof of such a statement will be provided {{elsewhere \cite{Prep}.}}
We thus end up with
\begin{align}
        R_{\text{se}} &= \Tr \left\lbrace 
   \mu^- \e^{\mathcal{L} t_3}\Big[\e^{\mathcal{L}t_2}\big[\mu^+ \e^{\mathcal{L}t_1}[\ket{g, 0}\bra{g, 0}\mu^-]\big]\mu^+\Big]
   \right\rbrace.
    \label{eq:stimulated_emission_response_function_splitted-2}
\end{align}
To evaluate $R_{\text{se}}$, we exploit
a Dyson-like expansion for the Lindblad equation 
characterizing the Markovian closure, i.e.,
\begin{align} \label{eq:lindTama}
    \frac{\d}{\d t} \rho(t) &= \mathcal{L}\big[\rho(t)\big]:= -\ii [H, \rho(t)] + \mathcal{D}\big[\rho(t)\big]
\end{align}
where the dissipator reads
\begin{align}
    \mathcal{D}[\rho] &:= \sum_{n=1}^{N} \Gamma_n \left( a_n \rho a_n^\dagger - \frac{1}{2} \lbrace a_n^\dagger a_n, \rho \rbrace \right).
\end{align}
The evolution map corresponding to such a Lindbladian can be in fact formally written as \cite{Hornberger2009}
\begin{eqnarray}
   && \e^{\mathcal{L} t}[\rho] = \e^{-\ii G t}\rho \e^{\ii G^\dag t} \label{eq:dyson} \\
   && +\sum_{j=1}^{\infty}\sum^N_{n_1,\ldots n_j = 1}\hspace{-0.3cm}\Gamma_{n_1}\ldots \Gamma_{n_j}
    \int_0^t d\tau_j \ldots \int_0^{\tau_2} d \tau_1
    \e^{-\ii G (t-\tau_j)}a_{n_j} \e^{-\ii G (\tau_j-\tau_{j-1})} \ldots a_{n_1}\e^{-\ii G \tau_1} \rho
    \e^{\ii G^\dag \tau_1}a^\dag_{n_1} \ldots \e^{\ii G^\dag (\tau_j-\tau_{j-1})}a^\dag_{n_j}\e^{\ii G^\dag (t-\tau_j)}
    \nonumber
\end{eqnarray}
where we introduced the non-Hermitian effective Hamiltonian
\begin{align} \label{eq:effHam}
    G &= H - \frac{\ii}{2} \sum_{n} \Gamma_n a_n^\dagger a_n.
\end{align}
Since $G \ket{g, 0} = a_{n} \ket{g, 0} = 0$, one easily sees that
\begin{align}
 \e^{\mathcal{L} t_1}\big[\ket{g,0}\bra{g, 0} \mu^-\big] =  \ket{g,0} \bra{g,0}\mu^-\e^{\ii G^\dag t_1},
    \label{eq:dynamical_map_pure_state_evolution}
\end{align}
which means that the jump contributions to the Dyson expansion (i.e., the terms of the form $a_{n_k}\sigma a^\dag_{n_k}$) are not relevant for single-time objects.

After the application of the second pulse at time $t_2$ we have
\begin{equation} \label{eq:t2}
    \mu^+ \ket{g,0}\bra{g, 0} \mu^- \e^{\ii G^\dagger t_1} = \rho_{t_1}.
\end{equation}
We observe that G acts non trivially on $\mu^+\ket{g,0} = \ket{E_b,0}$: the interaction between the system and the chain perturbs the state of the latter, and excitations start propagating along the chain. By applying the Dyson expansion \eref{eq:dyson} to the evolution determined by $\e^{\mathcal L t_2}$ on \eref{eq:t2} we have
\begin{eqnarray}
   && \e^{\mathcal{L} t_2}[\rho_{t_1}] = \e^{-\ii G t_2}\rho_{t_1} \e^{\ii G^\dag t_2} +\sum_{j=1}^{\infty}\sum^N_{n_1,\ldots n_j = 1}\Gamma_{n_1}\ldots \Gamma_{n_j}
    \int_0^{t_2} d\tau_j \ldots \int_0^{\tau_2} d \tau_1 \\
    &&\e^{-\ii G (t_2-\tau_j)}a_{n_j} \e^{-\ii G (\tau_j-\tau_{j-1})} \ldots a_{n_1}\e^{-\ii G \tau_1} \ket{E_b,0} \label{eq:partialt2} \\
    &&\bra{E_b,0} \e^{\ii G^\dagger t_1} \e^{\ii G^\dag \tau_1}a^\dag_{n_1} \ldots \e^{\ii G^\dag (\tau_j-\tau_{j-1})}a^\dag_{n_j}\e^{\ii G^\dag (t_2-\tau_j)}.
\end{eqnarray}
Let us focus on the term \eref{eq:partialt2}. Since $\tau_1<t_2$, the operators $a_{n_j}$ are expected to act trivially on $\e^{-i G \tau_1}\ket{E_b,0}$ as long as the oscillators in the closure are in the vacuum state. By a direct inspection of the chain parameters and considered the closure location $M=80$, the population of the closure oscillators starting from the state $\ket{g,0}$ can be estimated to be $T_M > 800 \unit[]{fs}$ (see, for example, the bottom-left frame of Fig.~\ref{fig:benchmark_tedopa_mc}). For the choice $t_2=500  \unit[]{fs}$ we can therefore safely neglect the jumps up to the application of the second pulse, so that 
\begin{equation}
   \e^{\mathcal{L}t_2}\big[\mu^+ \e^{\mathcal{L}t_1}[\ket{g, 0}\bra{g, 0}\mu^-]\big] \approx \e^{-\ii G t_2} \mu^+ \ket{g,0}\bra{g,0}\mu^- \e^{\ii (t_1+t_2)G^\dagger}.
\end{equation}
The effect of the jumps can no longer be neglected, in principle, once the oscillators in the closure get excited; this happens when the excitations travelling along the chain reach the closure. 
More precisely, for the considerations made above, the third order response function reads 
\begin{align}
        &R_{\text{se}} \approx 
      \bra{g, 0}\mu^-\e^{\ii G^\dag (t_1+t_2)}\mu^+
        \e^{\ii G^\dag t_3} \mu^- \e^{-\ii G (t_2+t_3)}\mu^+ \ket{g, 0}
   +\sum_{j=1}^{\infty}\sum^N_{n_1,\ldots n_j = 1}\Gamma_{n_1\ldots n_j}
    \int_0^{t_3} d\tau_j \ldots \int_0^{\tau_2} d \tau_1 \\
   &\times \bra{g,0} \mu^-
    \e^{\ii G^\dag (t_1+t_2+\tau_1)}
    a^\dag_{n_1} \ldots \e^{\ii G^\dag (\tau_{j}-\tau_{j-1})}a^\dag_{n_j} a_{n_j} 
    \e^{-\ii G (\tau_{j}-\tau_{j-1})} \ldots a_{n_1}\e^{-\ii G (\tau_1+t_2)} \mu^+\ket{g,0}.\nonumber
    \label{eq:dysonappl}
\end{align}
The jumps cannot be ignored when the last line of \eref{eq:dysonappl} differs from zero namely, following the same argument as for \eref{eq:partialt2}, for $t_3+t_2 \geq t_2+\tau_1 \gtrapprox T_M$. 

{For the computation of 2DES spectra discussed in the main text and in Section~\ref{sec:2d}, however, such error has always been negligible and certainly not discernible from the error generated by the other approximations related to the application of the Markovian closure itself (see Section~\ref{sec:TSO}). A clear example of this last kind of error is provided by the plots in last row of Fig.~\ref{fig:benchmark_tedopa_mc} showing the evolution of the average occupation number of the $80$-th site. In the setting of Fig.~\ref{fig:benchmark_tedopa_mc}, this is the site situated just before before the Markovian closure. We see that the error is generated, as expected, when the excitations start reaching the closure (around $0.8 \unit[]{ps}$). More precisely, the average occupation number of the 80-th site is slightly underestimated; this error is however related to the approximation introduced by the MC : as shown in Fig.~\ref{fig:fit_correlation_function}, the $N=6$ closure tends to slightly overdamp the exact two-time correlation function $C_\infty(t)$ defined in Eq.~\ref{eq:relation_asymptotic_semi_circle_cft}.
}

{The same Fig.~\ref{fig:benchmark_tedopa_mc}  allows to better understand why, for the specific case and parametrization of the MC, the evolution of pure states by means of an effective non-Hermitian dynamics suffices to reproduce the dynamics determined by the Lindblad master equation~\eref{eq:lindTama}. By inspecting  the average occupation number of the  $40$-th and of the $80$-th site (third and fourth row of Fig.~\ref{fig:benchmark_tedopa_mc}) it is in fact clear these chain sites, that are quite far from the system, are lowly occupied. The excitations generated by the interaction with the system and travelling along the chain arrive therefore to the closure at a low rate. On the other side, is designed to act as a ``perfect absorber'',  damping away incoming excitations at a rated that is, by  design, the same as the rate at wich the excitations arrive to the closure. A close look to the closure coefficients (see Tabs.~\ref{tab:summary_closure_parameters_N6}-\ref{tab:summary_closure_parameters_N10}), on the other side, allows to appreciate that there is a single auxiliary site in the closure having a damping rate ($\Gamma_n$) that is at least one order of magnitude larger than the other, and that is always of the same order as the transfer rate ($2 K$) from the last chain site (the $M$-th) to the closure. Most of the absorption takes therefore place at that site. The evolution of the closure is therefore happening in the single excitation subspace, where the average damping effect determined by the jump operators $a_n$ can be mimicked by the effective evolution determined by $G$. While we do not have a formal proof of this fact, all the numerical evidence points in this direction. This is indeed a most remarkable feature, allowing for the use of pure states not only in our 2DES related context.   
}

Summarizing, for the model at hand we can thus neglect the jump contributions, thus arriving at
\begin{align}
    R_{\text{se}} &\approx 
    \braket{\psi_{e,g}(t_2+t_1;t_{3}) | \psi_{e,g}(t_3+t_2;0)},
\end{align}
where we have defined
\begin{align}
    \ket{\psi_{e,g}(t;s)} &= \e^{-\ii G s} \hat{\mu}_- \e^{-\ii G t} \hat{\mu}_+ \ket{g, 0}.
\end{align}


In this work we focus on two-dimensional spectroscopy with respect to $t_1$ and $t_3$, i.e., we want to evaluate $R_{\text{se}}(t_1, t_2, t_3)$ for fixed $t_2 \equiv T_2$ and $t_1, t_3 \in [0,T_{13}]$.
Obviously we can not determine $R_{\text{se}}|_{t_2=T_2}$ on the continuous interval $[0,T_{13}] \times [0, T_{13}]$ but have to restrict to a discrete grid.
Given a certain number of sample points $N$ we can define
\begin{align}
    \mathcal{G} = \lbrace t_k := k \; \delta t \; | \; k = 1, \dots, N \rbrace
\end{align}
with a resolution in time of $\delta T := \frac{T_{13}}{N}$.
If we now want to evaluate $R_{\text{se}}|_{t_2 = T_2}$ on $\mathcal{G} \times \mathcal{G}$ we can proceed as follows.
First, we compute the evolution
\begin{align}\label{eq:one}
    \ket{\psi_{e}(t)} &:= \e^{-\ii G t}\hat{\mu}_{+}\ket{g, 0}
\end{align}
up to time $t \leq T_{13} + T_2$ and save the intermediate state for each $t = t_k + T_2$, $t_k \in \mathcal{G}$, to memory.
Second, for any $t = t_k + T_2$, $t_k \in \mathcal{G}$, we compute the evolution
\begin{align} \label{eq:two}
    \ket{\psi_{e,g}(s; t)} = \e^{-\ii G s} \hat{\mu}_{-}\ket{\psi_{e}(t)}    
\end{align}
up to time $s \leq T_{13}$.
For each $s_k \in \mathcal{G}$ load the previously computed state $\ket{\psi_{e}(s_k + T_2)}$ from memory, apply $\hat{\mu}_{-}$
\begin{align} 
    \ket{\psi_{e,g}(s_k + T_2; 0)} &:= \hat{\mu}_{-} \ket{\psi_{e}(s_k + T_2)},
\end{align}
and evaluate the overlap
\begin{align}
    R_{\text{se}}(t_k,T_2,s_k) &= \braket{\psi_{e,g}(t_k+T_2;s_k) | \psi_{e,g}(s_k + T_2; 0)}.
    \label{eq:third_order_response_function_grid}
\end{align}

\subsection{Simulation parameters, accuracy and computational performance}

\paragraph{Absorption spectra}

\begin{figure*}[hptb]
    \centering
    \includegraphics[width=\linewidth]{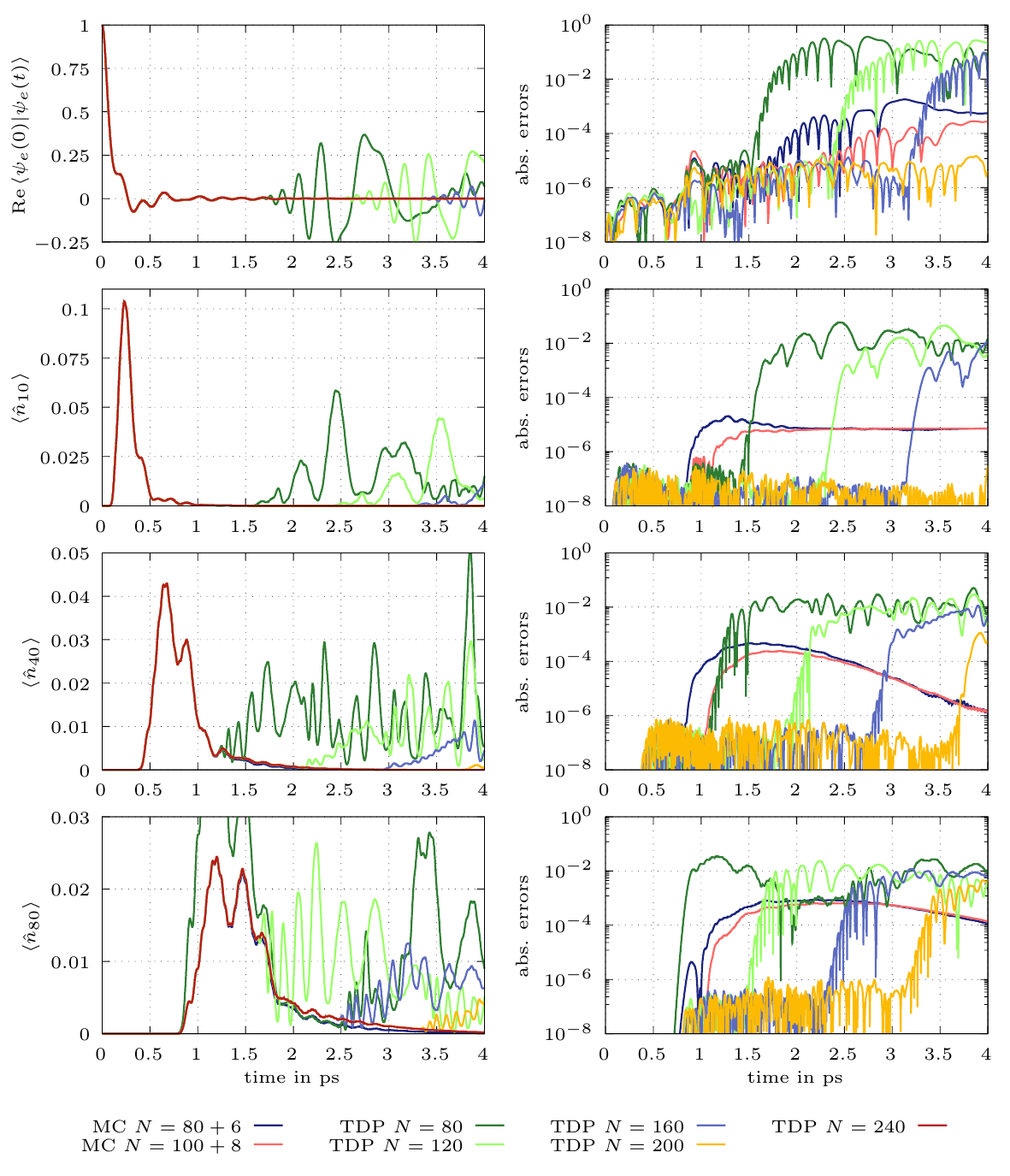}
    \caption{\textbf{Comparison of system and environmental observables within the computation of first-order response functions:} In the left column we depict the real part of the first-order response function defined in \cref{eq:first_order_response_function} as well a the average occupation on the $10^{\text{th}}$, $40^{\text{th}}$ and $80^{\text{th}}$ oscillator. These results are obtained by means of standard TEDOPA (TDP) with chain of different lengths ($N=80,120,160,200,240$), or by TEDOPA+MC (MC), with the closure made up of $6/8$ sites attached to the $80/100$-th site of the TEDOPA chain. In the right column we depict the corresponding absolute errors with respect to the standard TEDOPA solution with $N=240$.}
    \label{fig:benchmark_tedopa_mc}
\end{figure*}

In order to assess the computational performance and accuracy of TEDOPA+MC we considered the computation of the first-order response function required to evaluate absorption spectrum introduced in \cref{eq:absorption_line_shape}.
More precisely, the first-order response function is defined by 
\begin{align}
    R_{\text{abs}}(t) &:= \Tr{\Big\lbrace\hat{\mu}_{-} \mathcal{U}(t)\big[\hat{\mu}_{+} \ket{g,0} \bra{g,0}\big] \Big\rbrace} = \braket{\psi_{e}(0) | \psi_{e}(t)}
    \label{eq:first_order_response_function}
\end{align}
where
\begin{align}
    \ket{\psi_{e}(t)} &:= \mathcal{P}_t \big[\hat{\mu}_{-} \ket{g, 0}\big]
    \label{eq:first_order_response_function_evolved_state}
\end{align}
and the propagator is either unitary, $\mathcal{P}_t := \e^{-\ii \hat{H} t}$, for standard TEDOPA or non-unitary, $\mathcal{P}_t := \e^{-\ii \hat{G} t}$, for TEDOPA+MC.
For both approaches the time-evolved states are determined by means of the Time-Dependent Variational Principle (TDVP) method with single-site updates.
We found that a converged solution  \cite{conv}  is obtained using a local dimension of $D=6$ for all the oscillators in the chain and in the closure, a time step-width of $\delta t = \unit[0.5]{fs}$, and a MPS bond dimension of $\chi_{\ket{\psi}} = 8$.
The Hamiltonian corresponding to the TEDOPA chain admits an exact MPO representation of bond dimension $\chi_{\text{TDP}} =4$ whereas for the MC we need bond dimension $\chi_{\text{MC}}=6$ to account for the higher connectivity between oscillators.
The top row of \cref{fig:benchmark_tedopa_mc} shows a comparison of the results found with this configuration and variable extents of the environments.
Most importantly, it proves that TEDOPA+MC indeed yields accurate results over long simulation times without the necessity to increase the chain length further and further.

In addition, we monitored the population of several oscillators over time in order to emphasize the capability to access information about the environment within TEDOPA+MC.
Indeed, as \cref{fig:benchmark_tedopa_mc} indicates, TEDOPA+MC also yields accurate results for these environmental observables.
Even though the relative precision slightly degrades when looking at sites closer to the closure, it is at worst of the same order as the approximation error of the semi-circle spectral density, cf.~\cref{fig:fit_correlation_function}, and still provides reliable information about the primary environment.


As discussed in the main text, TEDOPA+MC suffers mainly two sources of error.
Firstly, the convergence of chain parameters is an asymptotic statement, i.e., any truncation after a finite number of sites induces an error. 
Secondly, the approximation of the semi-circle spectral density by means of an dissipative environment is error-prone.
As \cref{fig:benchmark_tedopa_mc} shows both of these errors can be systematically reduced by increasing the size of the primary environment or using closures with more constituents and thus higher accuracy, c.f.~\cref{fig:fit_correlation_function}.

\cref{fig:benchmark_tedopa_mc} clearly underlines that the MC is tailored to complement TEDOPA in order to be able to perform long-time simulations.
While for standard TEDOPA the maximal simulation time is in general a linear function of the chain length $N$, for TEDOPA+MC the maximal simulation time is independent of $N$ for a fixed desired precision.
This has two implications.
From the theory of matrix-product states we know that the memory cost scales as $\mathcal{O}(N D \chi^2 )$, i.e., linearly in $N$.
Hence, in contrast to standard TEDOPA the memory cost of TEDOPA+MC is bounded by a constant which is independent of the simulation time.
Furthermore, the computational time of the TDVP scheme with single-site updates scales linearly in $N$~\cite{lubich15,haege16}.
In total, we therefore expect the computational time to scale quadratically as a function of the simulation time for standard TEDOPA whereas for TEDOPA+MC we expect it to scale linearly.
This statement is further supported by Fig.3(e) of the main text. There we considered the precision of $\Re{\braket{\psi_e(0) | \psi_e(t)}}$ as a figure of merit and fixed $\epsilon = 10^{-3}$ as an upper bound on the error. Due to the lack of an analytical solution this error is taken with respect to a reference solution obtained via standard TEDOPA with $N=240$. While TEDOPA+MC with $N=80+6$ satisfies this criteria up to simulation time $\unit[4]{ps}$, for TEDOPA we steadily have to increase the number of oscillators in the chain. Hence, we first determined the maximal simulation time for $N \in \lbrace 20,40,60,\dots,200 \rbrace$ and then evaluated the corresponding computational time. 
\paragraph{2D spectra}

The computation of 2D spectra is slightly more involved.
In order to evaluate the third-order response function we have to compute the overlap of two non-unitarily evolved states on a two-dimensional grid with resolution $\delta T = \unit[1]{fs}$, $T_{13} = \unit[1]{ps}$ and $T_2 = \unit[0.5]{ps}$, c.f.~\cref{sec:computation_third_order_response} for further details.
The maximal simulation time for this grid is thus $\unit[2.5]{ps}$.
Again we use the TDVP evolution scheme with time step-width $\delta t = \unit[0.5]{fs}$, local dimension $D = 6$ for the first oscillator and $D^\prime = 4$ for the remaining oscillators, respectively.
While the bond dimension of the state is set to $\chi_{\ket{\psi}} = 8$, the MPO representation of the non-hermitian generator $\hat{G}$ has bond dimension $\chi_{TDP} = 4$ in the primary environment and $\chi_{MC} =6$ in the closure.

\bibliography{MarkovianClosure}